\documentclass[a4paper,11pt]{article}
\usepackage{amsmath,amssymb,mathrsfs,cite}
\usepackage{graphicx,enumerate,color,hyperref,ulem}

  \makeatletter
  \@addtoreset{equation}{section}
  \makeatother

\newcommand{\fr}[1]{\frac{1}{#1}}

\newcommand{\ord}[1]{{\mathcal O}(#1)}

\newcommand{\cA}{{\mathcal A}}

\newcommand{\cL}{{\mathcal L}}
\newcommand{\cP}{{\mathcal P}}

\newcommand{\nonum}{\nonumber\\ }

\newcommand{\sR}{{\sf R}}

\newcommand{\cout}[1]{}
\newcommand{\veps}{\varepsilon}
\newcommand{\bcP}{\bar{ {\cal P}}}
\newcommand{\zbar}{{\bar{z}}}

\begin{document}
\begin{titlepage}
\rightline{TTI-MATHPHYS-21}
\vskip 2cm
\vglue 2cm
\centerline{\LARGE \bf Dynamics of Myers-Perry black holes}
\vskip 0.5cm
\centerline{\LARGE \bf  with almost equal angular momenta in odd dimensions}

\vskip 1.6 cm
\centerline{\bf Ryotaku Suzuki and Shinya Tomizawa}
\vskip 0.5cm

\centerline{\small Mathematical Physics Laboratory, Toyota Technological Institute}
\centerline{\small 2-12-1 Hisakata, Tempaku-ku, Nagoya 468-8511, Japan}
\smallskip
\vskip 0.5cm
\centerline{\small\tt sryotaku@toyota-ti.ac.jp, tomizawa@toyota-ti.ac.jp}
\vskip 2cm



\centerline{\bf Abstract} \vskip 0.2cm \noindent
\noindent
We investigate the nonlinear dynamics of $D=2N+3$ Myers-Perry black holes with almost equal angular momenta, which have $N$ equal spins out of possible $N+1$ spins.
In particular, we study the ultraspinning instability and the fate of its nonlinear evolution using the large $D$ effective theory approach.
We find that every stationary phase can be mapped
 to the counterpart in the singly rotating phase within the leading order effective theory. From the known results of singly rotating solutions, we obtain the phase diagram of almost equally rotating black holes.
We also obtain a certain implication for the possible topology changing transition.

\end{titlepage}
\pagestyle{empty}
\small
\addtocontents{toc}{\protect\setcounter{tocdepth}{2}}
{
	\hypersetup{linkcolor=black,linktoc=all}
	\tableofcontents
}
\normalsize

\pagestyle{plain}
\setcounter{page}{1}

\section{Introduction}

\medskip

Higher-dimensional (more than four dimensions) black holes and other extended black objects have played  
a central role in the research of gravitational theories.
It is now evident that even within the vacuum Einstein theory, there is a much 
richer variety of black hole solutions in higher dimensions than in the four dimension since higher-dimensional black holes can have 
several rotations in multiple independent rotation planes
and the relative competition between the gravitational and centrifugal potentials is essentially different from four dimensions.
Such fascinating and profound nature is already found in the simplest rotating spherical solution derived by Myers and Perry~\cite{Myers:1986un}.
\medskip

In particular, in five-dimensional vacuum Einstein gravity, the first discovery of black ring solutions by Emparan and Reall~\cite{Emparan:2001wn} 
surprised many relativists because there are a rotating spherical hole and two rotating
rings with the same mass and angular momentum, providing the evidence against the uniqueness property of black holes in higher dimensions in contrast to the four dimension. 
By the development of the solution-generating methods which are applicable to the five-dimensional vacuum Einstein equations, various exact solutions of other black objects were subsequently found. 
Pomerasky and Sen'kov~\cite{Pomeransky:2006bd} first succeeded in the construction of the black ring solutions with two independent  rotations in the two orthogonal rotational planes.
Elvang and Figueras~\cite{Elvang:2007rd} constructed the exact stationary asymptotically flat vacuum solution describing black saturn, i.e., a spherical black hole surrounded by a black ring. 
Iguchi and Mishima~\cite{Iguchi:2007is} found the exact solutions of two concentric rotating black rings called black dirings, and Elvang, Rodriguez~\cite{Elvang:2007hs} and  Izumi~\cite{Izumi:2007qx} independently obtained the exact solutions of the dirings with spins around two independent planes describing two concentric orthogonal rotating black rings. 
Moreover, several authors~
\cite{Evslin:2008gx,Chen:2008fa,Tomizawa:2019acu,Lucietti:2020phh}
 tried to find the vacuum solutions of black lenses which have the horizon of lens spaces topologies but unfortunately all of these attempts failed, though such solutions were found as supersymmetric regular solutions in the five-dimensional minimal ungauged supergravity~
\cite{Kunduri:2014kja,Tomizawa:2016kjh}.
\medskip

In contrast to the stable Kerr family in $D=4$, Myers-Perry black holes have a multitude of instabilities at large enough angular momenta, which leads to richer dynamics. 
These so-called ultraspinning instabilities are caused by the deformation of the horizon shape due to the centrifugal forces~\cite{Emparan:2003sy}.
Since Myers-Perry black holes are rotating around the multiple axes, 
different spin configurations can lead to different dynamics. The existence of such instabilities are first confirmed by the linear perturbation in several setups~\cite{Dias:2010maa,Dias:2010eu,Dias:2011jg}.
The outcome of the ultraspinning instability are well understood in the singly rotating setup.
The numerical simulation revealed that this instability can end up with the fragmentation of the horizon accompanied by singular pinch-offs~\cite{Figueras:2017zwa} as observed in the black string simulation~\cite{Lehner:2010pn}. The zero modes of the instability also leads to the deformed stationary phases which eventually are connected to multi black rings or saturns through the the topology changing transition~\cite{Dias:2014cia,Emparan:2014pra}. 
On the other hand, such nonlinear phenomenon are not yet studied in more general configurations with multiple spins.

\medskip

In higher dimensions, the dynamical analysis of black holes has been relied heavily on the numerical calculation in most cases due to the lack of analyticity. The large dimension limit, or  {\it large $D$ limit}~\cite{Emparan:2013moa,Asnin:2007rw,Emparan:2020inr} provides a (semi-)analytic and systematic procedure  in the search of higher dimensional black holes at the cost of $1/D$-expansion. 
Remarkably, the black hole dynamics is reformulated into a certain effective theory on the horizon surface, which is often called {\it large $D$ effective theory}~\cite{Emparan:2015hwa,Bhattacharyya:2015dva,Bhattacharyya:2015fdk}.
\medskip


Moreover, the large $D$ limit offers another useful simplification for rotating black holes, such that the near horizon geometry at the leading order coincides with that of the static solution whose line elements are replaced by the local boosted frame~\cite{Emparan:2013xia},
so that the field equations become decoupled and integrable.
With this property, the large $D$ limit has been used for the study of the linear spectrum of the ultraspinning instability of equally rotating Myers-Perry black holes~\cite{Emparan:2014jca}, the dynamics of the singly rotating Myers-Perry black holes~\cite{Suzuki:2015iha}, as well as  
the construction of equally rotating black holes with the Maxwell charge~\cite{Tanabe:2016opw} and
Gauss-Bonnet correction~\cite{Suzuki:2022apk}.
 
\medskip

\begin{figure}[t]
\begin{center}
\includegraphics[width=15cm]{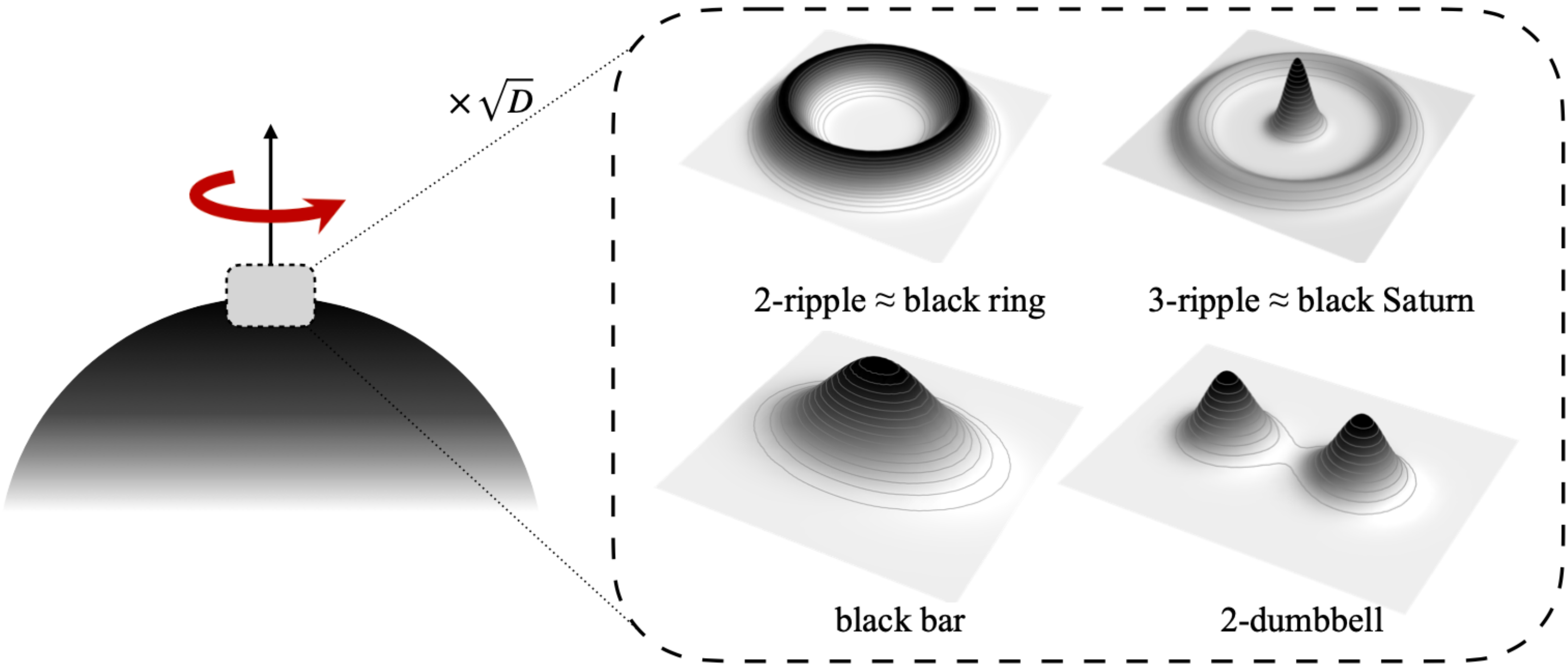}
\caption{Stationary phases of singly rotating black holes at large $D$ obtained by the blob approximation~\cite{Andrade:2018nsz,Andrade:2018rcx,Licht:2020odx,Suzuki:2020kpx}. The mass density profiles are given in the co-rotating cartesian coordinate.\label{fig:SR-phases}}
\end{center}
\end{figure}

In particular, the dynamics of singly rotating black holes at large $D$ has been investigated vigorously through the so-called {\it blob approximation}, in which the dynamics of a spherical black hole is identified as that of a Gaussian mass profile or {\it black blob} on the black brane magnifying the region around a polar axis of the black hole by $\sqrt{D}$~\cite{Andrade:2018nsz}.
The blob approximation was helpful in finding various (non)axisymmetric stationary phases at large $D$~\cite{Andrade:2018nsz,Andrade:2018rcx,Licht:2020odx,Suzuki:2020kpx} ( figure~\ref{fig:SR-phases} ). 
Although nonaxisymmetric horizons cannot remain stationary at finite $D$, they  would exist as long-lived intermediate states at large enough $D$ due to the $e^{-D}$-suppression of the radiation effect~\cite{Andrade:2019edf}.
In fact, the numerical simulation in $D=6,7$ actually provides the evidence of such long-lived intermediate states~\cite{Andrade:2020dgc}.
Beyond a single connected horizon, the blob approximation can also describe the collision of two separate black holes at large $D$~\cite{Andrade:2018yqu,Andrade:2020ilm,Luna:2022tgh}.
The similar approximation was also used to study the dynamics of AdS black holes~\cite{Emparan:2021ewh,Licht:2022rke,Emparan:2023dxm}.
 
 \medskip

In this article, we study the nonlinear dynamics derived from ultraspinning instabilities of equally rotating Myers-Perry black holes in $D=2N+3$ using the large $D$ limit. Focusing on the near polar dynamics, we obtain the dynamical effective theory of slightly broader case, i.e., $N$ equal angular momenta plus one different momentum out of $N+1$ angular momenta, which we call {\it almost equal rotation}. Surprisingly, we find that the large $D$ effective theory is equivalent to that of the singly rotating case.
Therefore, with the knowledge of singly rotating phases, the phase diagram is obtained for several sets of angular momenta.

\medskip
The rest of the article is organized as follows.
In section~\ref{sec:setup}, the metric ansatz and scaling assumptions are explained. The metric solutions expanded in $1/D$ are shown in section~\ref{sec:metricsol}. In section~\ref{sec:eft}, we present the large $D$ effective theory of almost equally rotating black holes.
The phase diagram of stationary solutions are studied in section~\ref{sec:stationary}.  We summarize and discuss the possible outcome in section~\ref{sec:end}.
We also attach an auxiliary Mathematica notebook that encloses the data for the metric solutions up to the next-to-leading order.

\section{Setup}\label{sec:setup}
In this article, we study rotating black holes with in $D=2N+3$ dimension, solving the vacuum Einstein equation
\begin{align}
R_{\mu\nu} = 0.
\end{align}
It is known that, if all $N+1$-spins are equal, $D=2N+3$ Myers-Perry black holes have the enhanced symmetry of $CP^N$ with the $S^1$-fibration~\cite{Kunduri:2006qa},
whose metric is given by
\begin{align}
ds^2 = -\frac{F(r)}{H(r)}dt^2+\frac{dr^2}{F(r)}+r^2 H(r)^2(\Phi+\Omega(r)dt)^2+r^2 d\Sigma^2_N,\quad \Phi := d\phi + \cA_N,\label{eq:ERMP-metric}
\end{align}
where $\cA_N$ and $d\Sigma_N^2$ are the K\"{a}hler potential and Fubini-Study metric on $CP^{N}$, respectively. The metric components are given by 
\begin{align}
F(r) = 1 - \frac{r_0^{2N}}{r^{2N}}+ \frac{a^2 r_0^{2N}}{r^{2N+2}},\quad
 H(r) = 1 + \frac{a^2 r^{2N}_0}{r^{2N+2}},\quad
 \Omega(r)=-\frac{a r_0^{2N}}{r^{2N+2} H(r)}.
\end{align}
The main advantage of the large $D$ limit is that the metric~(\ref{eq:ERMP-metric}) locally approaches to the large $D$ limit of the Schwarzschild metric~\cite{Emparan:2013xia,Emparan:2014jca}
\begin{align}
 ds^2 \simeq - \left(1-\fr{\sR}\right) (dt')^2 + \frac{d\sR^2}{D^2\sR(\sR-1)}+
r_0^2 (\Phi')^2+ r_0^2 d\Sigma_N^2,\quad \sR:= \frac{(r/r_0)^{2N}}{1-a^2}, \quad (D\sim N \gg 1) 
\end{align}
where the one forms $(dt',\Phi')$ is given by the local Lorentz boost of $(dt,\Phi)$,
\begin{align}
dt' := \cosh\alpha \, dt - \sinh\alpha \,\Phi,\quad
   \Phi' := \sinh\alpha\,  dt - \cosh\alpha \,\Phi,\quad \tanh\alpha:=a/r_0.
\end{align}
The same reduction is possible for other general cases of Myers-Perry black holes~\cite{Emparan:2013xia}.
Hence, the large $D$ limit of rotating black holes can be studied almost in parallel to the static case.

This metric admits the ultraspinning
instability both in axisymmetric and nonaxisymmetric modes with respect to the $\phi$ coordinate~\cite{Dias:2010eu,Emparan:2014jca}.
Here, we are interested in the end point of the instability at large $D$. 
To consider the nonlinear evolution of such instability, we must break the $CP^N$-symmetry. 
For this, it is convenient to decompose $CP^N$ in terms of $CP^{N-1}$ such that
\begin{align}
&d\Sigma_N^2 = d\theta^2 + \sin^2\theta \cos^2\theta \Psi^2 + \sin^2\theta d\Sigma^2,\nonum
&\cA_N = \sin^2\theta \Psi,\quad \Psi := d\psi + \cA,\label{eq:CPNtoCPNm1}
\end{align}
where $\cA$ and $d\Sigma^2$ are the K\"ahler potential and Fubini-Study metric on $CP^{N-1}$, respectively.

In the following analysis, we assume that the $U(1)$-symmetry for $\phi$ is broken in general, while the $U(1)$-symmetry for $\psi$ is always kept.
We also allow the black holes to have the angular velocity along $\psi$ that corresponds to $N$ spins out of $N+1$ spins\footnote{We present the Myers-Perry solution with such setup in Appendix \ref{sec:exactmp-n-spin}.},
which we call {\it almost equal} rotations. 
This is because, at the large $D$ limit, having $N$ equal spins and $N+1$ equal spins are approximately the same.


\paragraph{Blob approximation}
To treat the dynamical deformation of the spherical horizon correctly, we apply the blob approximation by introducing the near polar coordinate $z$ which zoom in around the $\psi$-axis at $\theta =\pi/2$, by $\sqrt{2N}$, such that
\begin{align}
 \theta := \frac{\pi}{2} - \frac{z}{\sqrt{2N}}.\label{eq:blobapprox}
\end{align}

In the singly rotating case, the similar approximation leads to the simpler formulation in which the dynamics of spherical black holes is obtained through the analysis of the Gaussian blob on the black $2$-brane~\cite{Andrade:2018nsz}.
Here we show that the almost equally rotating configuration admits the same approximation.

Let us demonstrate how the blob approximation improves a dynamical analysis at large $D$. For simplicity, we consider the scalar harmonics in a flat background
\begin{align}
\nabla^2 f =0,\label{eq:s2N+1-eq}
\end{align}
where the background metric is decomposed by using the $CP^{N-1}$ metric such that
\begin{align}
ds^2 = -dt^2+dr^2 + r^2\Phi^2+r^2\sin^2\theta \cos^2\theta \Psi^2 +r^2 d\theta^2 + 
r^2\sin^2\theta d\Sigma^2.
\end{align}
For the zero modes $f(r,\phi,\theta) = e^{im \phi} R(r) S(\theta)$, the angular part of eq.~(\ref{eq:s2N+1-eq}) becomes
\begin{align}
S''(\theta)+(N-1+N \cos (2 \theta )) \csc \theta  \sec \theta    S'(\theta)+ \left(\ell(\ell+2N)-m^2 \sec ^2\theta \right)S(\theta) =0,\label{eq:s2N+1-eq2}
\end{align}
where $\ell(\ell+2N)$ is the separation constant between $R(r)$ and $S(\theta)$.
Here we focus on the angular mode function $S(\theta)$.
The exact solution regular at $\theta = 0$ is obtained as
\begin{align}
S(\theta) = C\cos^m\theta \, {}_2 \- F_1\left(\frac{m-\ell}{2},\frac{m+\ell}{2}+N, N,\sin^2\theta\right).\label{eq:zeromode-exact}
\end{align}
For regularity at $\theta=\pi/2$, we must impose the quantization condition for $\ell$,
\begin{align}
 \ell = 2s+m,\quad s=0,1,2,\dots. \label{eq:zeromode-regularity}
\end{align}

In turn, by simply taking the limit $N\to\infty$, eq.~(\ref{eq:s2N+1-eq2}) reduces to the first order equation
\begin{align}
\cot \theta S'(\theta)+\ell S(\theta)=0,
\end{align}
which gives the leading solution in the $1/N$-expansion
\begin{align}
S(\theta) = C \cos^\ell \theta.\label{eq:zeromode-larged}
\end{align}
Clearly, this does not carry the information about the quantization condition~(\ref{eq:zeromode-regularity}).
It is straightforward to show that adding the $1/N$-correction does not improve the situation either.

The reason is because the limit $N\to\infty$ of $S(\theta)$ is not uniform in the entire range of $\theta \in [0,\pi/2]$, i.e., the $1/N$-expansion of the exact solution~(\ref{eq:zeromode-exact})
is consistent with eq.~(\ref{eq:zeromode-larged}) only in the range $|\theta-\pi/2| \gg 1/\sqrt{N}$\footnote{
We used the formula
\begin{align}
{}_2 F_1 (a,b+\lambda,\lambda,z) = (1-z)^{-a} \left(1-\frac{a b}{\lambda} \frac{z}{z-1}+\dots\right)\quad {\rm for} \quad \lambda \gg 1.
\end{align}}
\begin{align}
S(\theta) \simeq C \cos^\ell \theta \left( 1+ \ord{\tan^2\theta /N}\right).
\end{align}
Thus, the solution~(\ref{eq:zeromode-larged}) fails to capture the boundary behavior at $\theta = \pi/2$. One can also check that the zeros of the exact mode function~(\ref{eq:zeromode-exact}) with (\ref{eq:zeromode-regularity}) gather around $\theta-\pi/2 = {\cal O} (1/\sqrt{N})$ for $N \gg 1$.
In fact, the previous large $D$ analysis with $\theta$ had to impose the condition~(\ref{eq:zeromode-regularity}) as an extra condition to the effective theory~\cite{Emparan:2014jca,Tanabe:2016opw}.

On the other hand, the near polar coordinate $z$ in eq.~(\ref{eq:blobapprox}) properly resolves the regularity condition both at $\theta=0\ (z=\infty)$ and $\theta=\pi/2\  (z=0)$ at $N\to\infty$. With eq.~(\ref{eq:blobapprox}), eq.~(\ref{eq:s2N+1-eq2}) at $N\to\infty$ becomes
\begin{align}
z^2 \tilde{S}''(z) + z(1-2z^2)\tilde{S}'(z) + (2\ell z^2-m^2)\tilde{S}(z) =0,
\end{align}
where $\tilde{S}(z):=S(\pi/2-z/\sqrt{2N})$.
The solution regular at $z=0$ is given by
\begin{align}
 \tilde{S}(z) = C z^m\,  {}_1 \- F{}_1\left(\frac{m-\ell}{2},m+1,z^2\right).
\end{align}
It is easy to check that $\tilde{S}(z)$ diverges as $e^{z^2}$ for $z\to\infty$  unless eq.~(\ref{eq:zeromode-regularity}) is satisfied.

\paragraph{Metric Ansatz}

Having these in mind, we consider the following metric ansatz 
\begin{align}
 ds^2 = -A (e^{(0)})^2+2U e^{(0)} dr - 2C_i e^{(0)}e^{(i)} + H_{ij} e^{(i)}e^{(j)}+r^2 \sin^2 \theta d\Sigma^2, \label{eq:ansatz}
\end{align}
where $i,j=1,2,3$ and the boosted frame is given by
\begin{align}
&e^{(0)} := \gamma(dt-\Omega (d\phi + \sin^2\theta \Psi)),\quad
  e^{(1)} := \gamma(d\phi+\sin^2\theta \Psi-\Omega dt),\nonum
&e^{(2)} :=d\theta, \quad e^{(3)} :=  \sin\theta \cos \theta \Psi,\quad
\gamma :=\fr{\sqrt{1-\Omega^2}}.\label{eq:boostedframe}
\end{align}
In the following analysis, for convenience, we use 
\begin{align}
 n:=2N
\end{align}
as the large parameter and consider the expansion in $1/n$ instead of $1/D$ or $1/N$.
The near horizon coordinate are introduced by 
\begin{align}
\sR := (r/r_0)^n,  \label{eq:near-hcoord}
\end{align}
where we set $r_0=1$.
In the $z$-coordinate~(\ref{eq:blobapprox}), the boosted frame~(\ref{eq:boostedframe}) is approximated by
\begin{align}
&e^{(0)} = \gamma ( dt - \Omega (d\phi + \Psi))+\ord{n^{-1}} ,\quad e^{(1)} = \gamma ( d\phi+\Psi-\Omega dt)+\ord{n^{-1}},\nonum
&e^{(2)}  = -\frac{dz}{\sqrt{n}} ,\quad e^{(3)} = \frac{z}{\sqrt{n}} \left(\Psi+\ord{n^{-1}}\right).\label{eq:scaling-of-frame}
\end{align}
For later convenience, we also define the rescaled frame which remains $\ord{1}$ at $n\to\infty$,
\begin{align}
\hat{e}^{(0)} = e^{(0)},\quad \hat{e}^{(1)} := e^{(1)},\quad \hat{e}^{(2)} := -\sqrt{n} e^{(2)}\quad
 \hat{e}^{(3)} := \sqrt{n} e^{(3)}.
\end{align}
To obtain the reasonable solution at $n\to\infty$, we should scale each components of $C_i$ and $H_{ij}$ as
\begin{align}
  &C_1 = \fr{n} \hat{C}_1,\quad C_2 = -\fr{\sqrt{n}} \hat{C}_2,\quad C_3 = \fr{\sqrt{n}} \hat{C}_3,\nonum
&  H_{ii} = \hat{H}_{ii},\quad H_{23}=-\hat{H}_{23},\quad
  H_{12} = -\fr{\sqrt{n}} \hat{H}_{12},\quad H_{13} = \fr{\sqrt{n}}\hat{H}_{13}.
\end{align}
Then, the rescaled components are expanded in $1/n$ as functions of $(\sR,t,\phi,z)$,
\begin{align}
& A = \sum_{k=0}^\infty \frac{A^{(k)}}{n^k},\quad U = 1 + \fr{n} \sum_{k=0}^\infty \frac{U^{(k)}}{n^k},\quad  \hat{C}_i = \sum_{k=0}^\infty \frac{C_i^{(k)}}{n^k},\quad
\hat{H}_{ij} = \delta_{ij}+\fr{n} \sum_{k=0}^\infty \frac{H_{ij}^{(k)}}{n^k}.\label{eq:metricexpansion}
  \end{align}
To remain the asymptotic flatness, each variables should follow the asymtptotic boundary condition at $\sR\to\infty$ given in Appendix.~\ref{eq:asym-bc}.

\section{Metric solutions}\label{sec:metricsol}
Now, we solve the metric solution in the $1/n$-expansion~(\ref{eq:metricexpansion}) up to the next-to-leading order (NLO) by integrating the evolution equations with respect to $\sR$.
The constraint equations are imposed in the later section to obtain the effective equation.

\paragraph{Leading order}
The leading order solutions can be obtained as
\begin{subequations}\label{eq:LOsols}
\begin{align}
A^{(0)} = 1 - \frac{m(t,\phi,z)}{\sR},\quad C_i^{(0)} = \frac{p_i(t,\phi,z)}{\sR} -\frac{2\Omega}{1-\Omega^2} \log \sR\, \delta_{i1} ,
\end{align}
\begin{align}
H^{(0)}_{11}= \frac{2}{1-\Omega^2} \log \sR, \quad H^{(0)}_{ij} = 2\log \sR \delta_{ij}+\frac{p_i(t,\phi,z) p_j(t,\phi,z)}{m(t,\phi,z) \sR} \ {\rm for} \ (i,j)\neq(1,1)
\end{align}
and
\begin{align}
 U^{(0)} = -\frac{\Omega ^2 \log \sR}{1-\Omega ^2}-\frac{p_2(t,\phi ,z)^2+p_3(t,\phi ,z)^2}{2  m(t,\phi ,z)\sR},
\end{align}
\end{subequations}
where $m(t,\phi,z)$ and $p_i(t,\phi,z)$ are integration functions of $\sR$-integrals, which correspond to the mass density and momentum densities. Note that $m$ and $p_i$ are not arbitrary functions, but must be solutions of the effective equation that will be presented later.
The horizon is given by $\sR=m$ at the leading order.
The metric functions other than $A$ and $C_i$ are imposed the regularity at $\sR=m$ as well as the asymptotic boundary condition.

\paragraph{Next-to-leading order}
As usual in the large $D$ effective theory analysis, in the higher order, there are ambiguities in the solution of $A^{(k)}$ and $C^{(k)}_i$ that corresponds to the redefinition of $m$ and $p_i$. For simplicity, we set so that
\begin{align}
A^{(k)}(\sR=m)=0,\quad C^{(k)}(\sR=m)=0\quad (k > 0).
\end{align}
One should note that the former condition does not necessarily sets the event horizon at $\sR=m$ up to NLO. Later, we will see that the event horizon differs from $\sR=m$ at $\ord{1/n}$, when $m$ is not constant. As we will see later, the event horizon can differ from $\sR=m$ in general from $\ord{n^{-1}}$. The NLO solutions are given by
\begin{align}
 A^{(1)} =- \frac{2\Omega^2}{1-\Omega^2}\log \sR 
 +\frac{\left(-z \partial_z p_2 +(z^2-1)
   p_2 +\partial_\phi p_2 \right) \log (\sR/m)}{z \sR}
   -\frac{2 \Omega ^2 m  \log m }{(\Omega ^2-1)\sR},
\end{align}
and
\begin{align}
&C_1^{(1)} = \frac{\log (\sR/m)}{\sR} \left(\frac{ z \partial_z p_1p_2-\partial_\phi p_1 p_3+p_1 \left(z \partial_z p_2-(z^2-1)
 p_2-\partial_\phi p_3\right)}{z m }\right.\nonum
& \left.\hspace{5.3cm} + \ \frac{p_1 \left(\partial_\phi m p_3-z \partial_z m p_2\right)}{z m^2 }
\right)-\frac{2\Omega}{1-\Omega^2} \left(\log^2 \sR-\frac{m\log^2 m}{\sR}\right),\\
&C_2^{(1)} = \frac{\log (\sR/m)}{\sR} \left(-\frac{p_2   \left(\partial_\phi p_3-2 z \partial_z p_2\right)+p_3 \left(\partial_\phi p_2+p_3\right)+\left(z^2-1\right)
   p_2^2}{z m}\right.\nonum
&\left. \hspace{9cm}+ \  \frac{p_2 \left(\partial_\phi m p_3-z \partial_z m p_2\right)}{z m^2}+2 \gamma  \Omega  p_3\right),\\
&C_3^{(1)}=\frac{\log (\sR/m)}{\sR} \left(\frac{p_3
   \left(z \partial_z p_2-2 \partial_\phi p_3\right)+p_2 \left(z \partial_z p_3-\left(z^2-2\right) p_3\right)}{z  m}\right.\nonum
&   \left. \hspace{9cm}+\ \frac{p_3 \left(\partial_\phi m p_3-z \partial_z m p_2\right)}{z m^2}-2 \gamma  \Omega  p_2\right),
\end{align}
and
\begin{align}
 H_{11}^{(1)} =  -\frac{2\Omega^2 \log^2 \sR}{1-\Omega^2} + \frac{p_1^2}{\sR m}.
\end{align}
Because of the lengthy expressions, we will not show the detail, but other $H_{ij}^{(1)}$ and $U^{(1)}$ are also solved to evaluate thermodynamic variables up to NLO. We present the solutions up to NLO in the auxiliary Mathematica notebook.

\subsection{Local event horizon}
To keep track of the dynamical horizon, it is convenient to introduce a so-called local event horizon~\cite{Bhattacharyya:2008xc}, which is later used to evaluate the entropy and temperature. 

The position of the local event horizon $r=r_h(t,\phi,z)$ is defined as a null hypersurface $||dr-dr_h||^2=0$. In the boosted frame, the derivative of $r_h$ is written as
\begin{align}
dr_h = \partial_0 r_h e^{(0)} + \partial_i r_h e^{(i)} ,
\end{align}
where the dual basis is given by
\begin{align}
& \partial_0 := \gamma ( \partial_t + \Omega \partial_\phi),\quad
  \partial_1 := \gamma ( \partial_\phi+  \Omega \partial_t),\quad    \partial_2 := \partial_\theta = -\sqrt{n} \partial_z,\nonum
& \partial_3 := \csc\theta \sec\theta \partial_\psi - \tan \theta \partial_\phi
 = \frac{\sqrt{n}}{z}(\partial_\psi -\partial_\phi+\ord{n^{-1}}).
\end{align}
We also introduce the rescaled dual basis by
\begin{align}
\hat{\partial}_0:=\partial_0,\quad \hat{\partial_1}:=\partial_1,\quad \hat{\partial}_2 := - \fr{\sqrt{n}} \partial_2,\quad  \hat{\partial}_3 := \fr{\sqrt{n}} 
 \partial_3.
\end{align}
With the metric up to NLO, the null condition $||dr-dr_h||^2=0$ leads to
\begin{align}
\sR_h := r_h^n = m - \fr{n}\left(\frac{(p_2-\partial_z m)^2}{m}+\frac{(p_3-z^{-1}\partial_\phi m)^2}{m}-2 \gamma  z^2  ( \partial_t+\Omega \partial_\phi) m\right)+\ord{n^{-2}}.
\end{align}
The cross section with the $t={\rm const.}$ surface is then given by
\begin{align}
ds_H^2 = H_{ij} ( e^{(i)}-v^i e^{(0)})(e^{(j)}-v^j e^{(0)})+ r_h^2 \cos^2 (z/\sqrt{n}) 
 d\Sigma^2,\label{eq:dsH-boosted}
\end{align}
where the leading order solution~(\ref{eq:LOsols}) leads to
\begin{align}
 &v^1 = \fr{n} \hat{v}^1 =  \fr{n} \left(\frac{p_1-\hat{\partial}_1 m}{m}-2\gamma^2 \Omega \log m\right),\nonum
 &v^2 = -\fr{\sqrt{n}}\hat{v}^2= -\fr{\sqrt{n}}\frac{p_2 - \hat{\partial}_2 m}{m} , \nonum
 & v^3 = \fr{\sqrt{n}} \hat{v}^3= \fr{\sqrt{n}} \frac{p_3-\hat{\partial}_3 m}{m}.\label{eq:def-velocities-i}
\end{align}
In the coordinate basis, this is rewritten as
\begin{align}
ds_H^2 = \tilde{H}_{ab} ( dy^a - v^a dt)(dy^b-v^b dt) +r_H^2 \sin^2\theta d\Sigma^2, \label{eq:dsH-coord}
\end{align}
where $\tilde{H}_{ab}\ (a,b=\phi,z,\psi)$ is obtained from $H_{ij}$ with the transformation~(\ref{eq:frametrans}) and
\begin{align}
&v^\phi =  \Omega -v^\psi +\fr{n} (\gamma^{-2}\hat{v}_1+z^2 v^\psi)+\ord{n^{-2}}, \nonum
& v^z =  \gamma^{-1} \hat{v}^2+\ord{n^{-1}},
\nonum
& v^\psi = \frac{\hat{v}^3}{\gamma z}+\ord{n^{-1}}.
\end{align}

\section{Large $D$ effective  theory}\label{sec:eft}
Substituting the leading order solution~(\ref{eq:LOsols})
to the constraint equations on the $\sR={\rm const.}$ surface, we obtain the following effective equation
\begin{subequations}\label{eq:effeqz}
\begin{align}
&(\gamma   \partial_t +\gamma \Omega  \partial_\phi) m-\fr{z}\partial_\phi \left(p_3+\fr{z}\partial_\phi m\right)+ (\partial_z-z+z^{-1}) (p_2-\partial_z m)=0,
\label{eq:effeqz-0}\\
&(\gamma \partial_t +   \gamma  \Omega  \partial_\phi) p_1 - (\partial_z-z+z^{-1}) \left(\partial_z p_1-\frac{p_1p_2}{m}\right) - \fr{z}\partial_\phi \left( \fr{z} \partial_\phi p_1 +  \frac{ p_1 p_3 }{ m }\right)  \nonum
&  \quad  + \gamma^2\Omega\left(  (\partial_z-z+z^{-1}) (\partial_z m+p_2)+z^{-2}\partial_\phi^2 m-z^{-1}\partial_\phi p_3\right)    
\nonum
& \quad +2 \gamma   \left[z^{-1} \partial_\phi p_2+(\partial_z-z+z^{-1}) p_3 \right] +\gamma\partial_\phi m=0,
\label{eq:effeqz-1}\\
&(\gamma     \partial_t + \gamma  \Omega\partial_\phi) p_2- (\partial_z -z+z^{-1})\left(\partial_z p_2-\frac{p_2^2}{m}\right) -\fr{z}\partial_\phi \left( \fr{z}\partial_\phi p_2 + \frac{p_2p_3}{m}\right)
+\partial_z m    \left(\frac{2 \Omega ^2-1}{\Omega ^2-1}\right)\nonum
   &\quad -\frac{ 2\partial_\phi p_3 }{z^2}-\left(1-z^{-2}\right)  p_2-\frac{p_3^2}{zm}
   +2 \gamma  \Omega \left(p_3+z^{-1}\partial_\phi m\right)=0,
   \label{eq:effeqz-2}\\
&(\gamma     \partial_t +\gamma  \Omega\partial_\phi) p_3  - (\partial_z-z+z^{-1})\left(\partial_z p_3-\frac{p_2p_3}{m}\right)-\fr{z}\partial_\phi \left( \fr{z} \partial_\phi p_3 + \frac{p_3^2}{m}\right) 
-\fr{z}\partial_\phi m    \left(\frac{2 \Omega ^2-1}{\Omega ^2-1}
\right)\nonum
&\quad +\frac{2  \partial_\phi p_2}{z^2} +\left(1-z^{-2}\right)  p_3+\frac{  p_2   p_3}{z m }+2 \gamma  \Omega (\partial_z m-p_2) =0.
\label{eq:effeqz-3}
\end{align}
\end{subequations}

As discussed in ref.~\cite{Emparan:2015hwa}, these equations can be understood as an effective theory living in the overlap region ${\cal B}$ between the near-horizon and asymptotic regions, $1\ll \sR \ll e^n$ ( or $1\ll \log \sR \ll n$), where the metric on the ${\sf R}={\rm const.}$ surface becomes expanded both in $1/n$ and $1/\sR$ as
\begin{align}
 ds_{\cal B}^2 \simeq -dt^2+(d\phi+(1-z^2/n)(d\psi+\cA))^2+\fr{n}(dz^2+z^2 (d\psi+\cA)^2 )+(1-z^2/n) d\Sigma^2. \label{eq:metricB}
\end{align}

\paragraph{Spectrum around equally rotating Myers-Perry black holes}
The uniform solution corresponds to the equally rotating Myers-Perry black hole
\begin{align}
 m=1,\quad p_i=0.
\end{align}
One can see that the known spectrum are reproduced by considering the stationary perturbation
\begin{align}
 m = 1 + \veps e^{i k (\phi-\Omega t)} f_0(z),\quad p_i = \veps \delta p_i e^{i k (\phi-\Omega t)} f_i(z).
\end{align}
Plugging this into eqs.~(\ref{eq:effeqz}), we obtain the master equation for $f_0(z)$
\begin{align}
f_0''(z) -\left(z-\frac{1}{z}\right)
   f_0'(z)+   \left(\frac{1}{1-\Omega ^2}-\frac{k^2}{z^2}\right) f_0(z)=0,
\end{align}
and $f_i(z)$ are written in terms of $f_0(z)$.
By introducing the new variables
\begin{align}
 x := \frac{z^2}{2} ,\quad \tilde{f_0}(x) := z^{-k/2} f_0(z),
\end{align}
this reduces to the associated Laguerre equation
\begin{align}
   x \tilde{f_0}''(x) +(k+1-x) \tilde{f_0}'(x) + \fr{2}\left(\fr{1-\Omega^2}-k\right)\tilde{f_0}(x)=0.
\end{align}
Hence, the uniform solution has the normalizable stationary mode only if
\begin{align}
 \fr{2}\left(\frac{1}{1-\Omega^2}-k\right) = I ,\quad I=0,1,2,\dots
\end{align}
or
\begin{align}
 \Omega = \sqrt{1-\frac{1}{k+2I}}.
\end{align}
This is consistent with the perturbative analysis at large $D$~\cite{Emparan:2014jca,Tanabe:2016opw}.
The mode function is given by the associated Laguerre polynomial
\begin{align}
 f_0(z) = C z^k L^{k}_I\left(\frac{z^2}{2}\right).
\end{align}

\paragraph{Conservation form}
The physical meaning of the effective equation is clearer if we switch the variables from $(m,p_i)$ to $(m,\hat{v}^i)$.
Eq.~(\ref{eq:effeqz-0}) can be rewritten in the form of the mass conservation
\begin{align}
\hat{\partial}_0 \left( z e^{-\frac{z^2}{2}} m \right)+  \hat{\partial}_2 \left( z e^{-\frac{z^2}{2}} m \hat{v}_2\right)+ \hat{\partial}_3 \left( z e^{-\frac{z^2}{2}} m \hat{v}_3\right)  =0,
\end{align}
where the extra factor comes from the volume of the spacial cross section at large $D$
\begin{align}
 \sin\theta \cos\theta \times \sin^{n-2}\theta \simeq z e^{-\frac{z^2}{2}}.
\end{align}
We do not show the detail, but
one can also find that eqs.~(\ref{eq:effeqz-1}) and (\ref{eq:effeqz-3}) can be written in terms of the conservation of the Brown-York tensor in the direction of $\phi$ and $\psi$ as well. Note that eq.~(\ref{eq:effeqz-2}) cannot be written in the conservation, reflecting the fact that the asymptotic background is not symmetric in $z$ ( or $\theta$).

\subsection{Stationary solutions}
Now we show a stationary assumption reduces eq.~(\ref{eq:effeqz}) to a single master equation. First, we clarify the stationary condition for the effective theory.
From eq.~(\ref{eq:dsH-coord}), 
the horizon null generator $\xi$ is given by
\begin{align}
 \xi^\mu \partial_\mu = \partial_t + v^\phi \partial_\phi + v^z \partial_z + v^\psi \partial_\psi.
\end{align}
To obtain the stationary horizon, the horizon must stay at the same position along $\xi$
\begin{align}
 (\partial_t+v^a \partial_a) m=0. \label{eq:rigid-m}
\end{align}
As in ref.~\cite{Andrade:2018nsz}, we also require
that $\xi$ becomes the Killing vector of the metric~(\ref{eq:metricB}),
which imposes
\begin{align}
\partial_t v^a = 0, \label{eq:rigid-dtv}
\end{align}
and the shear free condition
\begin{align}
\sigma_{ab} := D_{(a} v_{b)} =0, \quad v_a:=\bar{h}_{ab} v^b,\label{eq:rigid-shearfree}
\end{align}
where $\bar{h}_{ab}$ is the spacial part of eq.~(\ref{eq:metricB}) and
$D_a$ are the covariant derivatives for $\bar{h}_{ab}$.

In the orthogonal frame of $\bar{h}_{ab}$, the shear tensor becomes
\begin{align}
 &\sigma_{11} \simeq \fr{\gamma^2 n} \partial_\phi \hat{v}_1,\quad 
 \sigma_{12} \simeq -\fr{2\sqrt{n}\gamma^2} (\partial_z \hat{v}^1 +\gamma \partial_\phi \hat{v}^2+2\gamma \hat{v}^3),\nonum
 & \sigma_{13} \simeq - \fr{2\sqrt{n} z \gamma^2}(\partial_\phi \hat{v}^1+2 z \gamma \hat{v}^2-z \gamma \partial_\phi \hat{v}^3) ,\nonum
&  \sigma_{22}\simeq\gamma^{-1}\partial_z \hat{v}_2,\quad
  \sigma_{23} \simeq \fr{2\gamma z}(-z \partial_z \hat{v}^3 + \hat{v}^3 + \partial_\phi \hat{v}^2),\quad
   \sigma_{33} \simeq \frac{\hat{v}^2-\partial_\phi \hat{v}^3}{\gamma z},\label{eq:sigmaij}
\end{align}
where $\sigma_{ij} := E_i^a E_j^b \sigma_{ab}$ with
\begin{align}
E^1_a dy^a = d\phi+(1-z^2/n+\dots)\Psi,\quad E^2_a dy^a = -\fr{n} dz,\quad E^3_a dy^a = \frac{z}{\sqrt{n}} \Psi+\dots.
\end{align}
The only solution that satisfies eqs.~(\ref{eq:rigid-dtv}) and (\ref{eq:rigid-shearfree}) and regular at $z=0$ is 
\begin{align}
\hat{v}^1 = -\gamma^2 z^2 \Omega_\psi + \Omega_1,\quad  \hat{v}^2 = 0,\quad \hat{v}^3 = \gamma  z \Omega_\psi,\label{eq:rigid-vi}
\end{align}
where $\Omega_1$ and $\Omega_\psi$ are the constants. 
This is equivalent to
\begin{align}
  v^\phi = \Omega-\Omega_\psi+\Omega_1/n ,\quad v^z = 0,\quad v^\psi = \Omega_\psi.\label{eq:rigid-vi}
\end{align}
Using the parameter shift $\Omega \to \Omega + \ord{1/n}$, we can always set $\Omega_1=0$ and the null generator becomes
\begin{align}
\xi^\mu \partial_\mu = \partial_t + (\Omega-\Omega_\psi) \partial_\phi + \Omega_\psi \partial_\psi. \label{eq:nullgenvector}
\end{align}
We should note that, although we started from the equally rotating ansatz, 
our setup also includes the states between $N+1$ equal spins for $\Omega_\psi=0$ and $N$ equal spins for $\Omega_\psi=\Omega$.

\medskip

Therefore, from the conditions~(\ref{eq:rigid-m}) and (\ref{eq:nullgenvector}),
$m$ must take the form of
\begin{align}
m =\exp \cP (\phi-(\Omega-\Omega_\psi) t, z), \label{eq:rigid-m2}
\end{align}
and $p_i$ are, then, expressed by $\cP$ through eqs.~(\ref{eq:def-velocities-i}) and (\ref{eq:rigid-vi}) as
  \begin{align}
 p_1 = (\hat{\partial}_1 \cP- \gamma^2 z^2 \Omega_\psi+2 \Omega \gamma^2 \cP)e^\cP ,\quad
 p_2 = \hat{\partial}_2 \cP  e^\cP,\quad p_3 = (\hat{\partial}_3 \cP + \Omega_\psi \gamma z) e^\cP.\label{eq:rigid-pi}
\end{align}
Plugging these into the effective equation~(\ref{eq:effeqz}),
we obtain a single master equation
\begin{align}
 \partial_z^2\cP + \left(z^{-1}-z\right) \partial_z \cP+\fr{z^2} \partial_\phi^2 \cP+ \fr{2} \left((\partial_z \cP)^2 + \fr{z^2} (\partial_\phi \cP)^2\right)+\gamma^2(\cP-\cP_0)+\frac{1}{2} \Omega_\psi ( \Omega_\psi-2\Omega) \gamma^2 z^2 =0,\label{eq:mastereq-P-gen}
\end{align}
where the parameter $\cP_0$ is the integration constant that fixes the horizon scale.
For simplicity, we set $\cP_0=0$.
A remarkable fact is that, with change of the variable
\begin{align}
 \cP =  -2(1-\Omega^2)+\frac{z^2}{2} + \bcP, \label{eq:Pbar-single}
\end{align}
eq.~(\ref{eq:mastereq-P-gen}) coincides with the master equation for the singly rotating case~\cite{Andrade:2018nsz}
\begin{align}
 \partial_\zbar^2 \bcP + \fr{\zbar} \partial_\zbar \bcP+\fr{\zbar^2} \partial_\phi^2 \bcP
 + \fr{2} \left((\partial_\zbar \bcP)^2 + \fr{\zbar^2} (\partial_\phi \bcP)^2\right)+\bcP+\fr{2}\omega_s^2 \zbar^2 =0,\label{eq:mastereq-single}
\end{align}
where
\begin{align}
 \zbar:= \gamma z ,\quad \omega_s := (\Omega_\psi-\Omega)\sqrt{1-\Omega^2}.\label{eq:vars-single}
\end{align}

Therefore, the results for the single rotating case~\cite{Andrade:2018nsz,Andrade:2018rcx,Andrade:2020ilm,Suzuki:2020kpx,Licht:2020odx} can be carried to the almost equally rotating case straightforwardly.
In the later section, we will see how the phase diagram is mapped as well.
One must note that this coincidence is not guaranteed up to the higher order in $1/n$.

\subsection{Thermodynamic variables}
\paragraph{ADM charges}
Reflecting the background symmetry in $t,\phi,\psi$, we have the conserved asymptotic charges of $M,J_\phi,J_\psi$, which are given in terms of Brown-York's quasi local tensor calculated by
\begin{align}
8 \pi G T^\mu{}_\nu := \lim_{\sR \to \infty} \sqrt{h} (K \delta^\mu{}_\nu - K^\mu{}_\nu) - 8 \pi G\tilde{T}^\mu{}_\nu,
\end{align}
where $\nabla_\mu$ and $K_{\mu\nu}$ is the covariant derivative and extrinsic curvature of the intrinsic metric $h_{\mu\nu}$ on the $\sR={\rm const}$-surface.
The regulator $\tilde{T}^\mu{}_\nu$ is determined so that $\nabla_\mu \tilde{T}^\mu{}_\nu=0$.
Corresponding to the background Killing vector $\partial_t, \partial_\phi, \partial_\psi$, these tensors satisfy the conservation law
\begin{align}
 \partial_\mu T^\mu{}_t =0,\quad
 \partial_\mu T^\mu{}_\phi =0,\quad
 \partial_\mu T^\mu{}_\psi =0.
\end{align}

The ADM mass is given by
\begin{align}
  M :=  
\frac{n\Omega_{n+1} }{16 \pi G}
\int \rho_M dV_{(z,\phi)},\quad \int dV_{(z,\phi)}:= \int_0^\infty dz \int_0^{2\pi}\frac{d\phi}{2\pi} ,
\end{align}
where the mass density up to NLO is calculated by the metric solution up to NLO as
\begin{align}
& \rho_M := -16 \pi G n T^t{}_{t} = \gamma^2 z e^{-\frac{z^2}{2}}  \left[  m + \fr{n} \left(
   \left(1+\frac{z^2}{3}-\frac{z^4}{12}\right)m - 2 \gamma^2 \Omega ^2 m \log m\right.\right.\nonum
&\hspace{4cm} \left.\left.     
   -\left(2-\Omega ^2+\log m \right) ( (\partial_z-z+z^{-1}) p_2-z^{-1} \partial_\phi p_3)\right.\right. \nonum
&\hspace{4cm}  + 2 \Omega   \left(p_1- \gamma \left(1-\Omega ^2\right) \partial_\phi m \right)
    - \gamma  \Omega ^2   (\partial_t+\Omega \partial_\phi) m \biggr)\biggr],
\end{align}
and we used the relation between the volume of $CP^{\frac{n}{2}-1}$ and that of $n$-sphere $\Omega_{n}$,
\begin{align}
\Omega_{n+1} = \frac{2\pi}{n} \Omega_{n-1} = \frac{(2\pi)^2}{n} \, {\rm vol}\left(CP^{\frac{n}{2}-1}\right).
\end{align}
Similarly, we define the angular momentum densities in $\phi$ and $\psi$ directions by
\begin{align}
 \rho_\phi = 16 \pi G n T^t{}_\phi,\quad \rho_\psi = 16 \pi G n T^t{}_\psi,
\end{align}
and the angular momenta are written by
\begin{align}
J_\phi = \frac{n\Omega_{n+1}}{16\pi G} \int \rho_\phi dV_{(z,\phi)},\quad
J_\psi = \frac{n\Omega_{n+1}}{16\pi G} \int \rho_\psi dV_{(z,\phi)}.
\end{align}
It turns out that $\rho_\phi$ and $\rho_\psi$ are proportional to $\rho_M$ at the leading order,
and hence we only show the difference at the sub-leading order
\begin{align}    
\rho_\phi = \Omega \rho_M + \frac{\delta \rho_\phi}{n},\quad \rho_\psi = \Omega \rho_M + \frac{\delta \rho_\psi}{n},
\end{align}
where
\begin{align}
&\delta \rho_\phi :=  z e^{-\frac{z^2}{2}} \left( \gamma^2\Omega  m+ p_1+  \Omega 
   (\partial_z -z+z^{-1})p_2- \gamma  \left(1-\Omega ^2\right) \partial_\phi m- \Omega  z^{-1} \partial_\phi p_3   \right),\\
&\delta \rho_\psi  := 
 z e^{-\frac{z^2}{2}} \left(
    \gamma^2 \Omega  \left(1-z^2\right) m + p_1+ z \gamma  p_3+ \Omega  (\partial_z-z+z^{-1}) p_2
   +  \gamma    \Omega ^2 \partial_\phi m
   - \Omega z^{-1} \partial_\phi p_3\right).
\end{align}
With the velocity fields~(\ref{eq:def-velocities-i}), this can be written as
\begin{align}
& \delta \rho_\phi = z e^{-\frac{z^2}{2}} m \left(\Omega \gamma^2+ \hat{v}^1+ 2\Omega \gamma^2 \log m\right)+ \partial_a (\cdots )^a,\\
& \delta \rho_\psi = z e^{-\frac{z^2}{2}}m \left(\hat{v}^1+ z \gamma \hat{v}^3 + 2\gamma^2 \Omega \log m+ \gamma^2 \Omega  (1-  z^2)\right)+ \partial_a (\cdots )^a,
\end{align}
where $\partial_a (\dots )^a$ denotes the spacial divergence.

\paragraph{Entropy}
The entropy of the dynamical horizon is calculated from the area of the local event horizon~(\ref{eq:dsH-coord}) as (see Appendix~\ref{sec:eventhorizon} for detail)
\begin{align}
S = \frac{ \Omega_{n+1}}{4G} \int \rho_S \, dV_{(z,\phi)},
\end{align}
where the entropy density $\rho_S$ is proportional to the linear combination of the mass and angular momentum and hence conserves at the leading order
\begin{align}
 \rho_S = \gamma (\rho_M - \Omega \rho_\phi) + \fr{n} \delta \rho_S.
\end{align}
In terms of the velocity fields~(\ref{eq:def-velocities-i}), the difference $\delta \rho_S$ is expressed as 
\begin{align}
\delta \rho_S =\gamma z e^{-\frac{z^2}{2}} \left[   \left(\gamma^2 \log m -1\right) m-\frac{m}{2}( (\hat{v}^2)^2+ (\hat{v}^3)^2)-\frac{  (\partial_z m) ^2}{2 m }-\frac{ (\partial_\phi m) ^2}{2 z^2 m }\right]+\partial_a (\dots)^a.
\end{align}
Therefore, the nonconserving subleading correction is given by the following functional
\begin{align}
&S_1 =  \frac{\Omega_{n+1}}{4G}\int \delta \rho_S\, dV_{(z,\phi)} \nonum
&=  \frac{\Omega_{n+1} \gamma}{4G} \int  z e^{-\frac{z^2}{2}} \left[   \left(\gamma^2 \log m -1\right) m-\frac{m}{2}( (\hat{v}^2) ^2+ (\hat{v}^3) ^2)-\frac{  (\partial_z m) ^2}{2 m }-\frac{ (\partial_\phi m) ^2}{2 z^2 m }\right]dV_{(z,\phi)}.\label{eq:def-S1}
\end{align}

\paragraph{Temperature}
The temperature is also defined  in terms of the surface gravity $T=\kappa/(2\pi)$ on the local event horizon if the horizon null generator~(\ref{eq:nullgenvector}) is the Killing vector (see Appendix \ref{sec:eventhorizon} for detail). Plugging the metric solution into eq.~(\ref{eq:def-temp}), we obtain
\begin{align}
T =& \frac{n}{4\pi \gamma} \left[1 + \frac{1}{n}\left(- 2 \gamma^2\Omega ^2-\gamma^2 \log m-\Omega  \hat{v}^1-\left(\partial_z+z^{-1}-z\right) \hat{v}^2-\frac{1}{2} (\hat{v}^2)^2-\frac{1}{2} (\hat{v}^3)^2+\frac{(\partial_z m)^2}{2 m^2}\right.\right.\nonum
&\left.\left.-\frac{\partial_z^2 m}{m}+\frac{\left(z^2-2 z \hat{v}^2-1\right) \partial_z m}{z m}+\frac{2 \hat{v}^3 \partial_\phi m}{z m}+\frac{(\partial_\phi m)^2}{2 z^2 m^2}+\frac{\partial_\phi \hat{v}^3}{z}-\frac{\partial_\phi^2 m}{z^2 m}-\frac{2
   \gamma  (\partial_t+\Omega \partial_\phi) m}{m} \right)\right].
\end{align}
By substituting eqs.~(\ref{eq:rigid-vi}) and (\ref{eq:rigid-m2}), the use of the master equation~(\ref{eq:mastereq-P-gen}) reduces this to the constant
\begin{align}
 T = \frac{n}{4\pi\gamma} \left(1 -\frac{2}{n}\gamma^2 \Omega^2 \right).
\end{align}
The extremal condition up to NLO is then given by
\begin{align}
 |\Omega| = 1 -\fr{2n}.\label{eq:extremalom}
\end{align}

\paragraph{First law and Smarr formula}
With the stationary conditions~(\ref{eq:rigid-vi}) and (\ref{eq:rigid-m2}), using eq.(\ref{eq:mastereq-P-gen}), one can show the first law up to $\ord{1/n}$ by taking the variation in $\cP$ and $\Omega_\psi$ independently
\begin{align}
 \delta M = T \delta S + \Omega_\phi \delta J_\phi + \Omega_\psi \delta J_\psi,
\end{align}
where $\Omega_\phi := \Omega-\Omega_\psi$.
The Smarr formula can also be shown up to $\ord{1/n}$
\begin{align}
 \frac{n+1}{n} M = TS+\Omega_\phi J_\phi + \Omega_\psi J_\psi. \label{eq:smarr}
\end{align}

\paragraph{Second law}
With the use of eqs.~(\ref{eq:effeqz}), the time derivative of eq.~(\ref{eq:def-S1}) leads to the second law
\begin{align}
&\partial_t S = \fr{n}\partial_t S_1\nonum
&=\frac{\Omega_{n+1}}{4G n} \int dV_{(z,\phi)} z e^{-\frac{z^2}{2}}\left(
 m  \left(z^{-1} \partial_\phi \hat{v}^2 - \partial_z \hat{v}^3 +z^{-1} \hat{v}^3 \right)^2+\frac{2 m}{z^2}  \left(\partial_\phi \hat{v}^3- \hat{v}^2 \right)^2+2  m  (\partial_z \hat{v}^2) ^2 \right) \geq 0.\label{eq:dS1dt}
\end{align}
Then, the entropy production ceases if 
\begin{align}
\hat{v}^2=0,\quad \hat{v}^3 = {\rm cosnt.} \times z.\quad \label{eq:v2v3rigid}
\end{align}
This is consistent with a part of the Killing condition~(\ref{eq:rigid-vi}). 
The contribution of $\hat{v}_1$ is suppressed to the subleading order in Eq.~(\ref{eq:dS1dt}), because of the scaling in eq.~(\ref{eq:scaling-of-frame}). Nevertheless, assuming the condition (\ref{eq:v2v3rigid}) in eq.~(\ref{eq:effeqz}), we obtain $m=\exp\cP(\phi-v^\phi t,z)$ with $v^\phi = {\rm const.}$ and the Laplace equation for $w(\phi-v^\phi t ,z):=\hat{v}^1(\phi-v^\phi t ,z)+\gamma z \hat{v}^3(z)$
\begin{align}
\partial_z \left(e^{\cP} z e^{-\frac{z^2}{2}} \partial_z w\right)
+\fr{z^2} \partial_\phi \left(e^{\cP} z e^{-\frac{z^2}{2}} \partial_\phi w\right) =0.
\end{align}
For the regular and normalizable solution both at $z=0$ and $z=\infty$, $w$ must be a constant as in ~(\ref{eq:rigid-vi}). 

It is less clear, but as in refs.~\cite{Andrade:2020ilm}, eq.~(\ref{eq:dS1dt}) can be written as the square of the viscosity tensor~(\ref{eq:sigmaij})
\begin{align}
 \partial_t S = \frac{\Omega_{n+1}}{4Gn} \int z e^\frac{z^2}{2} 2\gamma^2 m  h^{ab} h^{cd}\sigma_{ac} \sigma_{bd} dV_{(z,\phi)}, \label{eq:dS1dt-viscos}
\end{align}
where $h_{ab}$ is the spacial part of eq.~(\ref{eq:metricB}).
Here we note that only $\sigma_{22},\sigma_{23},\sigma_{33}$ contribute to the leading order in eq.~(\ref{eq:dS1dt}).

\subsection{Reduction to the singly rotating phase}

All the above thermodynamic quantities can be mapped to those of singly rotating black holes through eqs.~(\ref{eq:Pbar-single}) and (\ref{eq:vars-single}). First, we define normalized thermodynamical quantities by the mass scale to eliminate the scaling degree of freedom. The angular momenta normalized by the mass scale become
\begin{align}
& j_{\phi} := \frac{16\pi G J_{\phi}}{(n+2)\Omega_{n+1}r_M^{n+1}} = \tilde{\Omega} + \fr{n} \delta j_{\phi},\quad j_{\psi} := \frac{16\pi G J_{\psi}}{n\Omega_{n+1}r_M^{n+1}}= \tilde{\Omega} + \fr{n} \delta j_{\psi},
\end{align}
where $\tilde{\Omega}:=\Omega r_M$ and the mass scale is given by
\begin{align}
 r_M := \left(\frac{16\pi G M}{(n+1)\Omega_{n+1}}\right)^\fr{n}\simeq 1+\fr{n}\log \left(\int \rho_M dV_{(z,\phi)}\right).\label{eq:mass-scale}
\end{align}
$\delta j_\phi$ is given by
\begin{align}
 \delta j_\phi = \frac{\int z e^{-\frac{z^2}{2}} (m \gamma^{-2} \hat{v}_1+2\Omega m \log m) dV_{(z,\phi)}}{\int z e^{-\frac{z^2}{2}} m \, dV_{(z,\phi)}}- 2\Omega \log \left(\int z e^{-\frac{z^2}{2}} \gamma^2 m dV_{(z,\phi)} \right).\label{eq:def-delta-phi}
\end{align}
Instead of $\delta j_\psi$, it is more convenient to see the difference between $j_\psi$ and $j_\phi$ since it vanishes for the equally rotating Myers-Perry ($j_\phi=j_\psi$),
\begin{align}
\Delta  j := \delta j_\psi -\delta j_\phi =n \left( j_\psi- j_\phi\right)  &= \frac{\int z^2 e^{-\frac{z^2}{2}}( \gamma^{-1} m \hat{v}^3+ \Omega(2-z^2)  m)   dV_{(z,\phi)}}{\int  z e^{-\frac{z^2}{2}} m dV_{(z,\phi)}}
. \label{eq:def-delta-psi}
\end{align}
Note that we keep using $\Omega$ instead of $\tilde{\Omega}$
 in eqs.~(\ref{eq:def-delta-phi}) and (\ref{eq:def-delta-psi}) as the difference only comes in the higher order~(\ref{eq:mass-scale}).
 
 With the stationary conditions~(\ref{eq:rigid-vi}) and (\ref{eq:rigid-m2}), or equivalently (\ref{eq:rigid-pi}), the normalized angular momenta~(\ref{eq:def-delta-phi}) and (\ref{eq:def-delta-psi}) become
\begin{align}
&\delta j_\phi =- \Omega_\psi \frac{\int z^3 e^{-\frac{z^2}{2}}  e^\cP dV_{(z,\phi)}}{\int  z e^{-\frac{z^2}{2}} e^\cP dV_{(z,\phi)}} + 2 \Omega \frac{\int z \cP e^{-\frac{z^2}{2}}  e^\cP  dV_{(z,\phi)}}{\int  z e^{-\frac{z^2}{2}} e^\cP dV_{(z,\phi)}} - 2\Omega \log \left(\int  z e^{-\frac{z^2}{2}} e^\cP dV_{(z,\phi)}\right),\label{eq:deltajphi-rigid}\\
&\Delta j = 2 \Omega - (\Omega-\Omega_\psi)\frac{\int z^3 e^{-\frac{z^2}{2}}  e^\cP  dV_{(z,\phi)}}{\int  z e^{-\frac{z^2}{2}} e^\cP dV_{(z,\phi)}}.
\label{deltaj-rigid}
\end{align}
We show that these have simpler expressions using the singly rotating variables calculated in refs.~\cite{Andrade:2018nsz,Licht:2020odx,Suzuki:2020kpx}.
In particular, with eqs.~(\ref{eq:Pbar-single}) and (\ref{eq:vars-single}), eq.~(\ref{deltaj-rigid}) can be expressed as
\begin{align}
 \Delta j = 2 \Omega + \sqrt{1-\Omega^2} j_s,\label{eq:dj-js}
\end{align}
where $j_s=j_s(\omega_s)$ is the normalized angular momentum for the singly rotating black holes given by the solution to eq.~(\ref{eq:mastereq-single}) as
\begin{align}
 j_s :=\frac{\int \omega_s \bar{z}^3 e^{\bcP} dV_{(\bar{z},\phi)}}{ \int \bar{z} e^{\bcP} dV_{(\bar{z},\phi)}},
\end{align}
and the angular velocity of the single rotation $\omega_s$ is given by $\Omega$ and $\Omega_\psi$ through eq.~(\ref{eq:vars-single}). Note that, by definition, $j_s$ is an odd function of $\omega_s$.
Eq.~(\ref{eq:deltajphi-rigid}) is also simplified as
\begin{align}
 \delta j_\phi &=  -\sqrt{1-\Omega^2} j_s + 2 \Omega (h_s - \log \mu_s)\nonum
& = - \Delta j+ 2 \Omega (h_s +1- \log \mu_s) ,\label{eq:res-deltajphi}
\end{align}
where we used eq.~(\ref{eq:dj-js}) and $h_s=h_s(\omega_s)$ and $\mu_s=\mu_s(\omega_s)$ are calculated from the singly rotating solution
\begin{align}
 h_s :=\frac{\int 
 \bar{z} \bcP e^{\bcP}dV_{(\bar{z},\phi)}
 }{\int\bar{z}e^{\bcP}dV_{(\bar{z},\phi)} },
 \quad \mu_s := \int \bar{z} e^{\bcP}  dV_{(\bar{z},\phi)}.\label{eq:def-hs}
  \end{align}
  Once we specify the corresponding singly rotating family of the phase $j_s=j_s(\omega_s)$, $\Omega_\psi$ is determined as the functions of $(\Omega,\Delta j)$ through
eqs.~(\ref{eq:vars-single}) and (\ref{eq:dj-js}).
The entropy~(\ref{eq:def-S1}) is normalized in the same way and expressed in terms of the singly rotating variables
\begin{align}
s := \frac{4GS}{\Omega_{n+1} r_M^{n+1} } =\fr{\tilde{\gamma}} \left[1+\frac{\gamma^2}{n}\left(-\Omega \delta j_\phi +(2\Omega-\Delta j)\Omega-\omega_s j_s - \log \mu_s + 1-3\Omega^2\right)\right]. \label{eq:def-norm-s}
\end{align}

\paragraph{Myers-Perry phases}
As shown in ref.~\cite{Andrade:2018nsz}, eq.~(\ref{eq:mastereq-single}) admits the singly rotating Myers-Perry solution
\begin{align}
 \bcP = \frac{2}{1+\bar{a}^2}\left(1-\frac{\bar{z}^2}{4}\right),
\end{align}
which leads to $j_s := 2\bar{a}$ and
\begin{align}
\omega_s = \frac{2 j_s}{4+j_s^2},\quad 
 \mu_s = \left(1+\frac{j_s^2}{4}\right) e^\frac{8}{4+j_s^2},\quad h_s = 1-\omega_s j_s.\label{eq:mpvar-single}
 \end{align}
 One can easily check that the corresponding almost equally rotating phase is actually the Myers-Perry phase of a given $(\Omega,\Delta j)$.
To confirm this, let us compare the corresponding phase with the large $D$ limit of the exact Myers-Perry solution with almost equal angular momenta given in Appendix~\ref{sec:exactmp-n-spin}.
The first equation of eq.~(\ref{eq:mpvar-single}) is inversely solved as
\begin{align}
 j_s = \frac{1\pm \sqrt{1-4\omega_s^2}}{\omega_s}.
\end{align}
Then, eqs.~(\ref{eq:vars-single}) and (\ref{eq:dj-js}) reproduce eq.~(\ref{eq:exactmp-dj1}) at the leading order.
The normalized entropy~(\ref{eq:def-norm-s}) is also consistent with eq.(\ref{eq:exactmp-dj}) at the leading order.

\paragraph{Entropy difference}
Since the entropy~(\ref{eq:def-norm-s}) is insensitive to the horizon dynamics at the leading order, it is convenient to see only the difference from a reference phase.
For this, we compare with the Myers-Perry solution with the same $(j_\phi,j_\psi)$.
Since $j_\phi \simeq j_\psi \simeq \Omega$, $\Omega$ is also the same at the leading order. Nevertheless, one should recall that we also have the degree of freedom to be adjusted that changes $\Omega$ in $\ord{n^{-1}}$ as in eq.~(\ref{eq:rigid-vi}). Thus, having the same $j_\phi$ does not necessarily means the same $\delta j_\phi$, instead requires 
\begin{align}
j_\phi = \Omega + \fr{n}( \Omega_1 + \Omega \log \mu_s + \delta j_\phi) 
= \Omega + \fr{n} (\Omega_1^{\rm MP} + \Omega \log \mu_s^{\rm MP} + \delta j_\phi^{\rm MP}),\label{eq:cond-jphi}
\end{align}
where the quantities with ${\rm MP}$ denotes those for the Myers-Perry phase. This condition determines the difference in $\Omega_1$.

Then, the entropy difference from the Myers-Perry phase becomes
\begin{align}
 s-s_{\rm MP} =\frac{\gamma}{n}\left( -\omega_s j_s+\omega_s^{\rm MP} j_s^{\rm MP}-\log (\mu_s/\mu_s^{\rm MP}) \right),
\end{align}
where eq.~(\ref{eq:cond-jphi}) is used with the fact that
\begin{align}
\fr{\tilde{\gamma}}-\fr{\tilde{\gamma}^{\rm MP}} \simeq \frac{\gamma}{n}\Omega (-\Omega_1+\Omega_1^{\rm MP}-\Omega \log (\mu_s/\mu_s^{\rm MP}) ).
\end{align}
To avoid the divergent behavior near the extremality, we define the entropy difference as
\begin{align}
\Delta s := -\omega_s j_s+\omega_s^{\rm MP} j_s^{\rm MP}-\log (\mu_s/\mu_s^{\rm MP} ).
\label{eq:def-deltas}
\end{align}
Since eq.~(\ref{eq:dj-js}) leads to
\begin{align}
 j_s^{\rm MP} = \frac{\Delta j-2\Omega}{\sqrt{1-\Omega^2}} = j_s,
\end{align}
eq.~(\ref{eq:mpvar-single}) means
\begin{align}
\omega_s^{\rm MP} = \frac{2 j_s}{4+j_s^2} ,\quad  \mu_s^{\rm MP} = \left(1+\frac{j_s^2}{4}\right) e^\frac{8}{4+j_s^2}.
\end{align}
Thus, eq.~(\ref{eq:def-deltas}) is simplified to
\begin{align}
 \Delta s = 2-\omega_s j_s- \log \mu_s + \log\left(1+\frac{j_s^2}{4}\right).\label{eq:res-deltas}
\end{align}
Interestingly, this is identical to the entropy difference in the singly rotating case, in which case the comparison should be made with the singly rotating Myers-Perry phase\footnote{It has not been shown explicitly, but one can easily obtain from the entropy formula in ref.~\cite{Andrade:2020ilm}.}.
This suggests the thermodynamics for given $(\Omega, \Delta j)$ is identical with the corresponding 
singly rotating phase with $(\omega_s, j_s)$.
Hence, in the following, we do not distinguish $\Delta s$ for the almost equally rotaing case and that for the singly rotating case.

\section{Stationary phases}\label{sec:stationary}

\subsection{Stationary phases in the singly rotating case : a review}
\begin{figure}[t]
\begin{center}
\includegraphics[width=12cm]{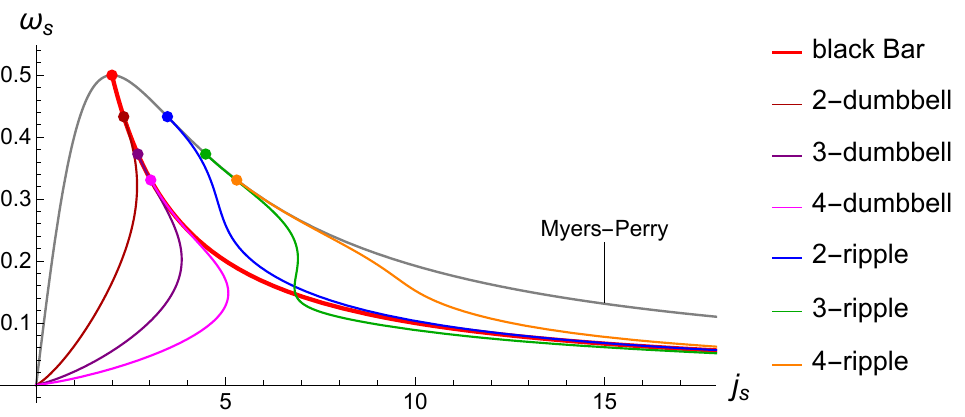}
\caption{Stationary phases of the singly rotating black hole in $(j_s,\omega_s)$~\cite{Licht:2020odx}.\label{fig:js-ws-single}}
\end{center}
\end{figure}

Since the stationary phase in the almost equally rotating setup is mapped to the singly rotating phase, 
we first revisit the stationary phases in the singly rotating setup found in refs.~\cite{Andrade:2018nsz,Licht:2020odx,Suzuki:2020kpx}.
In figure~\ref{fig:js-ws-single}, the major stationary phases in the singly rotating case are shown:
\begin{itemize}
\item Myers-Perry black holes
\item Black ripples : Axisymmetric deformed branches of the Myers-Perry phase bifurcating from axisymmetric zero modes at $j_s=2\sqrt{2k-1},\ (k=2,3,4,\dots)$. Here we call the corresponding branches as $k$-{\it ripple}. At the large deformation and large angular momentum, the horizon tends to be fragmented into several parts of ring shapes with/without a central Gaussian blob for even/odd branches, respectively. At finite $D$, they would eventually cause the topology-change to black multi-rings/saturns. 
\item Black bar : A nonaxisymmetric branch bifurcates from the zero mode of the bar mode instability at $j_s=2$~\cite{Andrade:2018nsz}. 
At large deformation and large angular momentum, the profile tends to be infinitely elongated in one direction and becomes unstable to the Gregory-Laflamme type instability.
\item Black dumbbells : Deformed black bar branches bifurcate from the zero mode instabilities of the black bar at $j_s=2k/\sqrt{2k-1},\ (k=2,3,4,\dots)$, which we call {\it $k$-dumbbell} for each $k$. At large deformation, both the angular momentum and angular velocity tend to be zero and the shape is
fragmented into a multiple array of orbiting black blobs. 
\end{itemize}
Typical profiles of these phases are shown in figure~\ref{fig:SR-phases}.
A remarkable feature of the large $D$ effective theory is that it admits the nonaxisymmetric stationary solutions, which cannot be stationary at finite $D$ due to the radiation.
This is because the effects of the gravitational wave emission is suppressed nonperturvative in $D$, i.e. $\sim e^{-D}$, and hence negligible in the $1/D$-expansion~\cite{Andrade:2019edf}. One may doubt that such solutions are artifacts only at $D=\infty$. However, one may also interpret them as long-living intermediate states due to the low radiation rate at large enough $D$. In fact, the black hole-black hole collision simulation in $D=6,7$ captures relatively long-living black bar and dumbbell states~\cite{Andrade:2020dgc}.
\begin{figure}[t]
\begin{center}
\includegraphics[width=8.5cm]{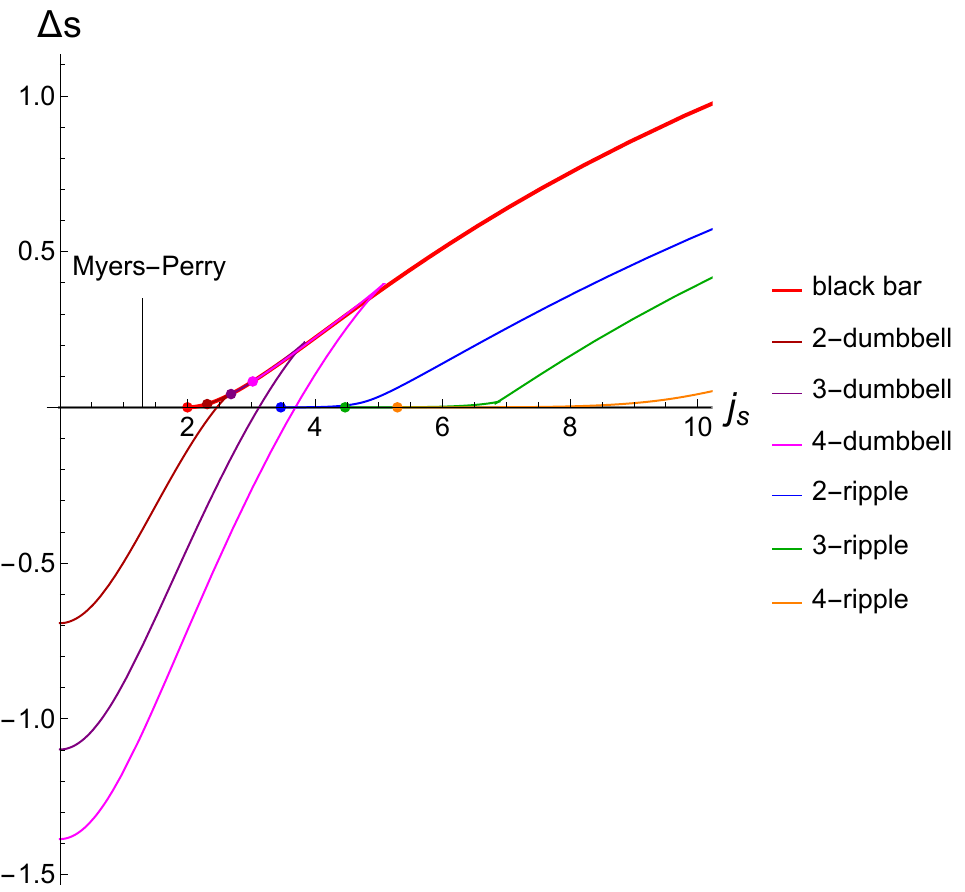}
\begin{minipage}[b]{7cm}
\includegraphics[width=7cm]{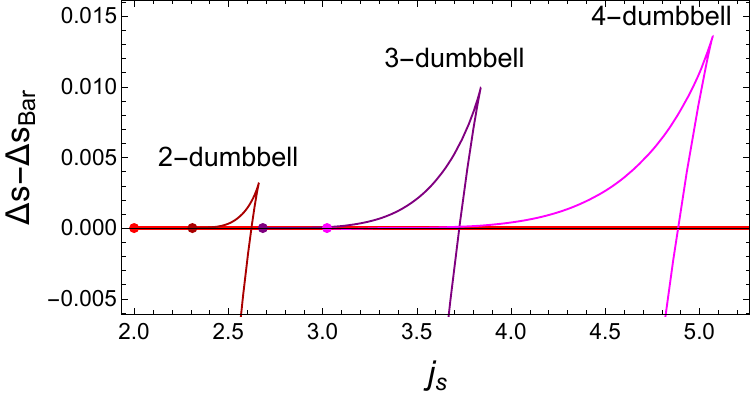}\\\\
\includegraphics[width=7cm]{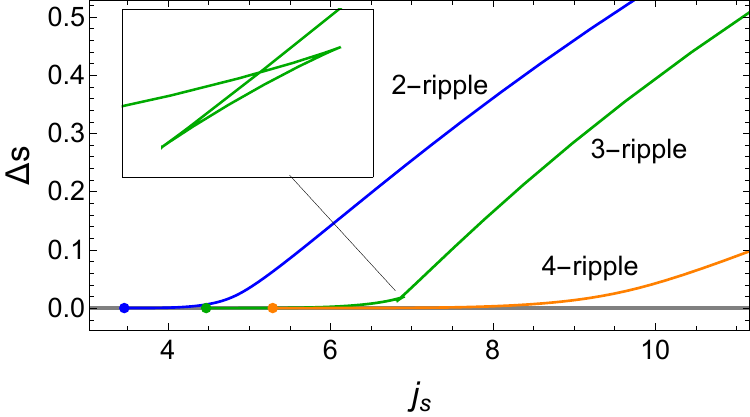}
\end{minipage}
\caption{Entropy of singly rotating black holes compared with Myers-Perry black holes of the same mass and angular momentum, which are partially shown in ref.~\cite{Andrade:2020ilm}.
The close-ups for dumbbells and ripples are shown in the right panel. \label{fig:entropy-single}}
\end{center}
\end{figure}

In figure~\ref{fig:entropy-single}, we show the entropy of the above solutions given by eq.~(\ref{eq:res-deltas}).
Beyond the bar mode at $j_s=2$, the most stable state is the black bar until the first zero mode takes place on the black bar at $j_s = 4/\sqrt{3}$.  Then, $2$-dumbbell branch becomes the most stable state beyond $j_s=4/\sqrt{3}$ until it reaches to a turning point $j_{s,{\rm crit}}\approx 2.662$. The numerical simulation in the large $D$ effective theory indicates that the black hole collision results in 
the most entropically favored phase for the corresponding total angular momentum, where the collision always ends up in the fragmented profile for $j_s > j_{s,{\rm crit}}$~\cite{Andrade:2020ilm}.

\begin{figure}[t]
\begin{center}
\includegraphics[width=8.5cm]{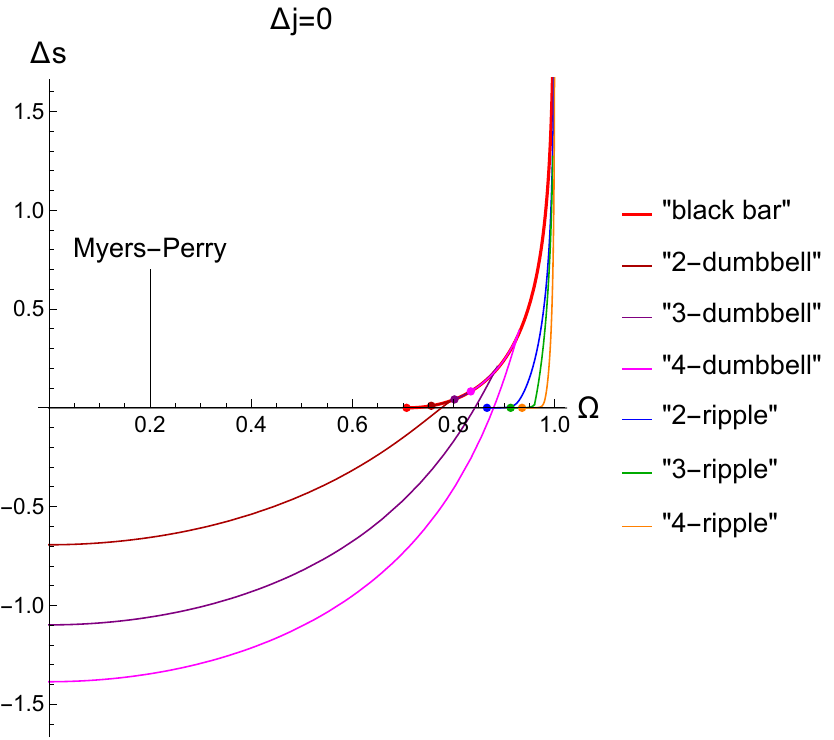}
\begin{minipage}[b]{7cm}
\includegraphics[width=7cm]{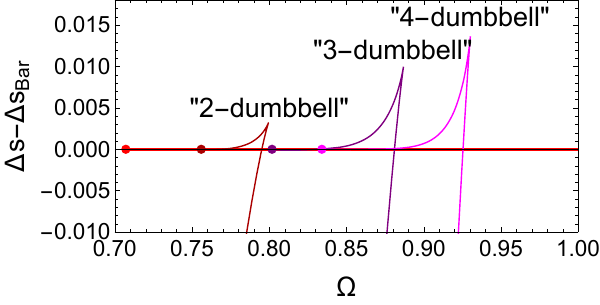}\\\\
\includegraphics[width=7cm]{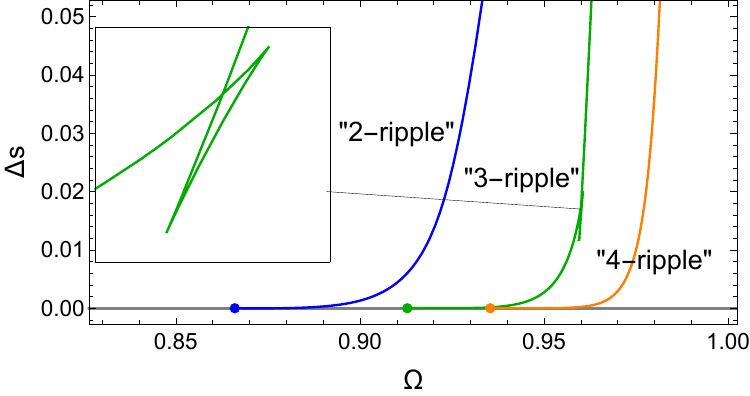}
\end{minipage}
\caption{Phase diagram of almost equally rotating black holes with $\Delta j=0$. Each phases are labeled after the name of singly rotating counterparts. \label{fig:dsplots_dj0}}
\end{center}
\end{figure}

\subsection{Phase diagram}
Finally, we present the phase diagram of almost equally rotating black holes in figure~\ref{fig:dsplots_dj0}. Each phases are labelled by the name of corresponding singly rotating phases with double quotation, e.g., black bar $\to$ ``black bar". 
From singly rotating phases in figure~\ref{fig:js-ws-single}, each solutions with $(\omega_s, j_s)$ are mapped to $(\Omega,\Omega_\psi)$ through eqs.~(\ref{eq:dj-js}) and (\ref{eq:vars-single}) for given $\Delta j$. The corresponding solutions have the same entropy difference~(\ref{eq:res-deltas}).
Since the parameters has the symmetry of simultaneous flip of the sign
\begin{align}
(\omega_s,j_s,\Omega,\Omega_\psi,\Delta j) \to (-\omega_s,-j_s,-\Omega,-\Omega_\psi,-\Delta j),
\end{align}
we can always choose $\Delta j \geq 0$ and instead consider the range $\Omega \in (-1,1)$.

An important point is that eq.~(\ref{eq:dj-js}) sets the different parameter region for $j_s$ for $\Omega \in (-1,1)$ depending on the value of $\Delta j$:
\begin{enumerate}
\item{$0\leq \Delta j < 2$: } 
$j_s$ monotonically decreases from $j_s=\infty\, (\Omega=-1)$ to $j_s=-\infty\, (\Omega=1)$. Eq.~(\ref{eq:dj-js}) is solved as
\begin{align}
\Omega = \frac{2 \Delta j-j_s \sqrt{4-\Delta j^2+j_s^2}}{4+j_s^2}.\label{eq:js-om-case1}
\end{align}
\item{$\Delta j = 2$: }
$j_s$ monotonically decreases from $j_s=\infty\, (\Omega=-1)$ to $j_s=0\, (\Omega=1)$. $\Omega=\Omega(j_s)$ is also given by eq.~(\ref{eq:js-om-case1}).

\item{$\Delta j > 2$: } 
$j_s$ reaches to the minimum at $\Omega = 2/\Delta j$ and goes to $\infty$ both at $\Omega \to \pm 1$, that is, the value of $j_s$ is bounded below
\begin{align}
 j_s \geq \sqrt{\Delta j^2-4}.
\end{align}
In this case, $\Omega$ is written as the multi-valued function of $j_s$
\begin{align}
\Omega_{\pm} =\frac{2 \Delta j\pm j_s \sqrt{4-\Delta j^2+j_s^2}}{4+j_s^2},
\end{align}
where $-1<\Omega_-<2/\Delta j<\Omega_+<1$.
\end{enumerate}

\begin{figure}[t]
\begin{center}
\includegraphics[width=8cm]{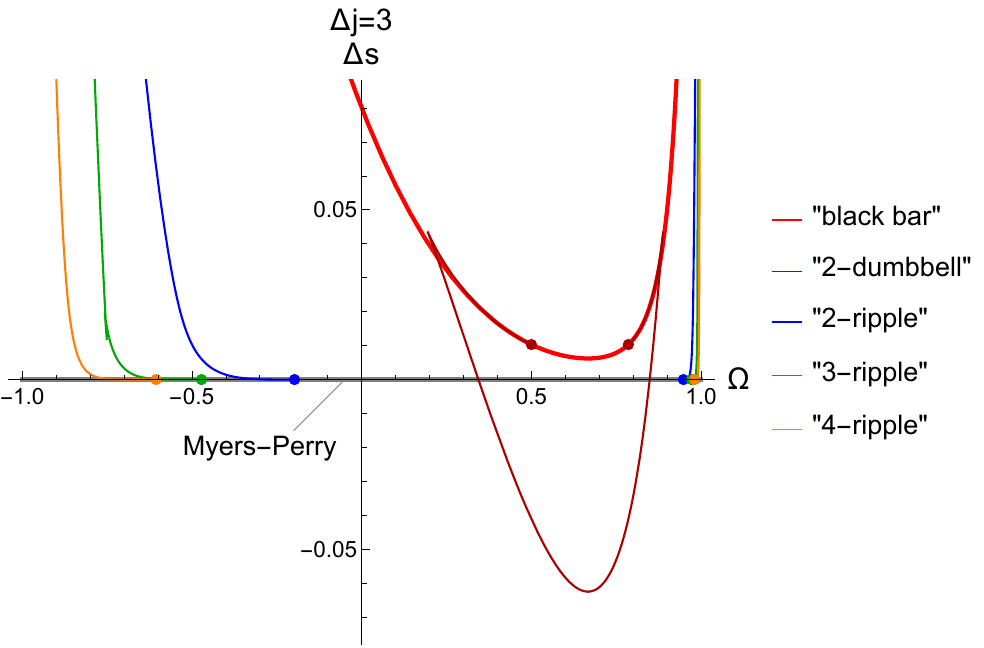}
\includegraphics[width=7cm]{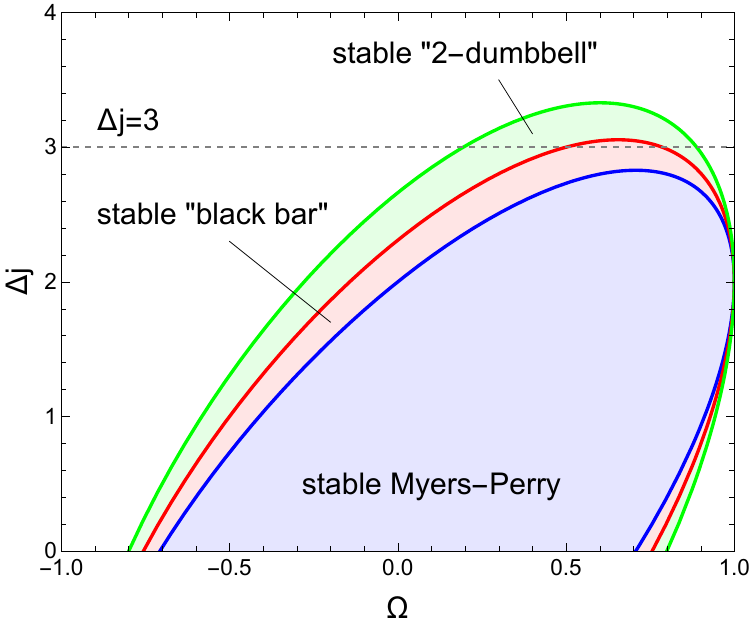}
\caption{{\bf Left panel:} The phase diagram for $\Delta j=3$. Higher ``dumbbells" are omitted for visibility. {\bf Right panel:} The stable phases for given $(\Omega,\Delta j)$. The stable phase is Myers-Perry in the blue region ($j_s \leq 2$), ``black bar" in the red region ($2\geq j_s\leq 4/\sqrt{3}$) and ``$2$-dumbbell" in the green region ($4/\sqrt{3} \leq j_s \leq 2.662 $). Beyond the green region, none of these three are stable. \label{fig:dsplot-dj-b}}
\end{center}
\end{figure}

In the third case, particularly, the lower bound of $j_s$ largely affects the appearance of the phase diagram.
For example, if $\Delta j > 2 \sqrt{2}$, with which $j_s$ is always above the bar mode threshold, the Myers-Perry solution cannot be stable and the black bar branch no longer bifurcates from the Myers-Perry phase (left panel of figure~\ref{fig:dsplot-dj-b}). We will not show all cases but the same phenomenon occurs for higher other branches as well at larger $\Delta j$.
We also present the most preferable phase for given $(\Omega,\Delta j)$ in the right panel of figure~\ref{fig:dsplot-dj-b}. 

\begin{figure}[t]
\begin{center}
\includegraphics[width=16cm]{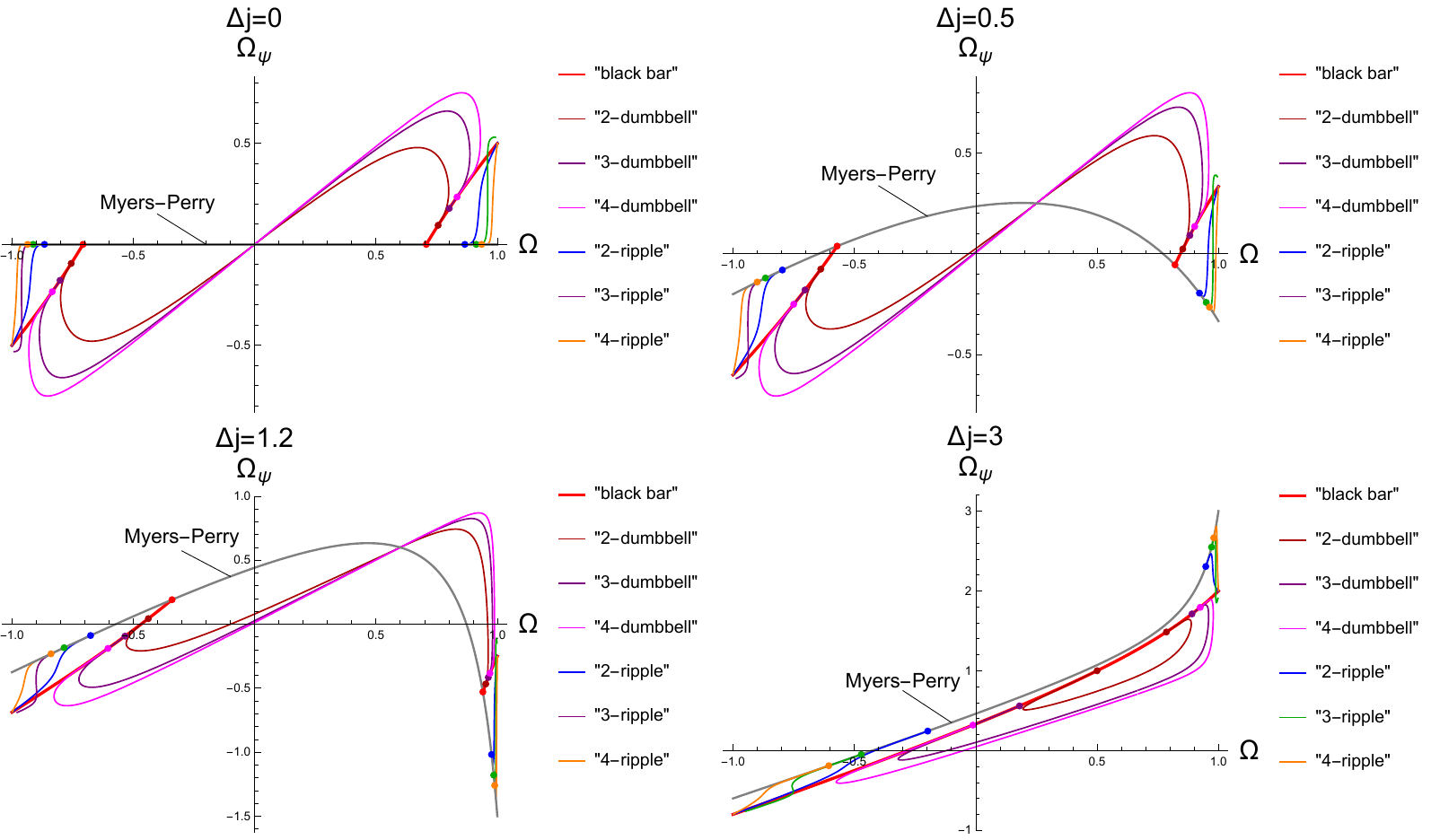}
\caption{$\Omega_\psi$ for each phases with several values of $\Delta j$.\label{fig:ompsi}}
\end{center}
\end{figure}

$\Omega_\psi$ for each phases is plotted in figure~\ref{fig:ompsi}.
It is interesting to note that even if we set $\Delta j=0$, the deformed phases have nonzero $\Omega_\psi$. This can be understood that the conservation of $\Delta j$ in eq.~(\ref{deltaj-rigid})
makes the difference in $\Omega_\psi$ due to the difference in the moment of inertia.

\section{Discussion}\label{sec:end}
In this article, we have investigated the dynamics of rotating black holes with almost equal angular momenta, which have $N$ equal spins out of $N+1$ spins, using the large $D$ effective theory approach. We have first studied the nonlinear dynamics involving the ultraspinning instability in the equally rotating case, which is found to be captured by the simple effective theory under the proper scaling ansatz at large $D$. 
We have found that the effective theory of almost equally rotating black holes reduces to that of singly rotating black holes in the stationary setup, and thermodynamical variables are also mapped to singly rotating counterparts. In particular, we have found that the phase diagram for stationary solutions shares the common microcanonical structure in both setups.
 This implies that all the instabilities and deformed branches found in the singly rotating setup have their counterparts in the almost equally rotating setup. 
\medskip

At finite $D$, the singly rotating setup admits axisymmetrically deformed phases, which are called bumpy black holes or ripples~\cite{Dias:2014cia,Emparan:2014pra}.
It is expected that those branches are connected to branches of different topologies such as black rings or black saturns through the topology changing transition.
In the blob approximation at large $D$, this topology changing transition takes place in a smooth and more unclear way.
For example, the black ripples at the large deformation limit tend to be isolated ring-like Gaussian blobs with or without a central blob connected via thin black branes~\cite{Licht:2020odx,Suzuki:2020kpx} (figure~\ref{fig:SR-phases}). 
Since the $1/D$-expansion breaks down if the mass density $m$ becomes $m\sim e^{-D}$, such highly-deformed phases no longer describe the blob approximation of a single deformed black hole, but should rather describe black rings or saturns. This occurs for $j_s \sim \sqrt{D}$ and $\omega_s \sim 1/\sqrt{D}$~\cite{Suzuki:2020kpx}. Then, one can see from eq.~(\ref{eq:dj-js}) that this corresponds to $\Omega = 1 -\ord{1/D}$ in the almost equally rotating setup. 
However, the big difference from the singly rotating case is that $\Omega$ has the upper bound due to the extremality which is also given by $\Omega_{\rm ext} = 1-\ord{1/D}$~(\ref{eq:extremalom}).
Therefore, in the almost equally rotating setup, the topology changing transition would not occur
if the phase reaches its extremality before the transition.
To confirm this, one has to examine higher order corrections in $1/D$.

\medskip

If the topology changing transition really ocurrs, another question is, what is the topology after the transition? Can they admit novel horizon topologies?
Since the current ansatz has a twisted structure on $CP^{N-1}$, the horizon topology after the pinch-off is less clear than in the singly rotating ansatz. For this, the soap bubble analysis would hint the shape of the horizon~\cite{Emparan:2015hwa}. This will be postponed to the future work.
\medskip

It is straightforward to extend the analysis to more general cases with the cosmological constant, Maxwell field~\cite{Tanabe:2016opw},
or with the Gauss-Bonnet correction~\cite{Suzuki:2022apk}.
It would be worth to examine whether the similar large $D$ correspondence between single rotating black holes and almost equal rotating black holes holds in such general cases.

\medskip

In the AdS background, even in finite $D$, the equally rotating setup is known to admit nonaxisymmetric AdS black holes with only one Killing vector, called {\it black resonators}~\cite{Dias:2011at,Dias:2015rxy,Ishii:2018oms,Ishii:2019wfs,Ishii:2020muv,Ishii:2021xmn}. They are surrounded by the matter clouds or gravitational radiation which are maintained by the superradiant instability. One may pursue the possibility of such resonator-type solutions at large $D$, by
applying the similar large $D$ analysis in the AdS background. 
However, finding the equilibrium condition for the radiation would be difficult task at large $D$, since the effect of the radiation is suppressed by $e^{-D}$, and furthermore the radiating sector belongs to the higher frequency modes of $\omega \sim D$~\cite{Emparan:2014aba}, which are not described by the effective theory approach.
\medskip

By adding a Kaluza-Klein circle to equally rotating black holes, one can obtain equally rotating black strings.
The large $D$ analysis in this setup will lead to black strings  with nontrivial helical Killing vector as seen in  $D=6$~\cite{Dias:2022mde,Dias:2022str}.
This would also be an intriguing direction.

\section*{Acknowledgement}
R.S. was supported by JSPS KAKENHI Grant Number JP18K13541. 
S.T. was supported by JSPS KAKENHI Grant Number 21K03560.

\appendix

\section{ADM decomposition with $CP^{N-1}$ reduction}
In the section~\ref{sec:metricsol}, we solve the following reduced equation instead of the Einstein equation itself to save the computational resource\footnote{The computation of the $1/D$-expansion on the Mathematica takes considerably longer times as the dimension of the non-symmetric subspace increases.}.
We consider the decomposition into the radial direction and others,
\begin{align}
 ds^2 = \alpha^2 d\rho^2 + \bar{g}_{\bar{\mu}\bar{\nu}}
 (d\bar{x}^{\bar{\mu}}+\bar{\beta}^{\bar{\mu}} d\rho)( d\bar{x}^{\bar{\nu}}+\bar{\beta}^{\bar{\nu}} d\rho),
\end{align}
with which the Einstein equation is decomposed to the evolution equation
\begin{align}
\fr{\alpha} \partial_\rho \bar{K}^{\bar{\mu}}{}_{\bar{\nu}} = \bar{R}^{\bar{\mu}}{}_{\bar{\nu}}-\bar{K}\bar{K} ^{\bar{\mu}}{}_{\bar{\nu}} + \fr{\alpha} \cL_{\bar{\beta}} \bar{K} ^{\bar{\mu}}{}_{\bar{\nu}}- \fr{\alpha}\bar{\nabla}^{\bar{\mu}} \bar{\nabla}_{\bar{\nu}}\alpha,\label{eq:adm-evol0}
\end{align}
the scalar constraint
\begin{align}
\bar{K}^2 - \bar{K}^{\bar{\mu}}{}_{\bar{\nu}} \bar{K}^{\bar{\nu}}{}_{\bar{\mu}} - R =0,\label{eq:adm-scalarc0}
\end{align}
and the vector constraint
\begin{align}
\bar{\nabla}_{\bar{\nu}} \bar{K}^{\bar{\nu}}{}_\mu - \bar{\nabla}_{\bar{\mu}} \bar{K} =0,\label{eq:adm-vectorc0}
\end{align}
where $\bar{K}^{\bar{\mu}}{}_{\bar{\nu}}$ and $\bar{R}^{\bar{\mu}}{}_{\bar{\nu}}$ are the extrinsic and intrinsic curvature for $\bar{g}_{\bar{\mu}\bar{\nu}}$, respectively.

We assume the intrinsic geometry has the isometry of a $S^1$ fiber over $CP^{N-1}$
\begin{align}
 \bar{g}_{\bar{\mu}\bar{\nu}}d\bar{x}^{\bar{\mu}} d\bar{x}^{\bar{\nu}} = g_{\mu\nu}(x)\xi^\mu \xi^\nu +h(x)^2 \gamma_{ab} dy^a dy^b,\quad \xi^\mu = dx^\mu+\delta^\mu{}_\psi \cA_a dy^a,
\end{align}
in which $\gamma_{ab}$ and $\cA_a$ are the Fubini-Study metric and K\"ahler potential of $CP^{N-1}$ and $\psi$ is the fiber coordinate.
The shift vector is given by
\begin{align}
 \bar{\beta}_{\bar{\mu}} d\bar{x}^{\bar{\mu}} = \beta_\mu dx^\mu + \beta_\psi \cA_a dy^a,\quad
 \bar{\beta}^{\bar{\mu}} \partial_{\bar{\mu}}  = \beta^\mu \partial_\mu.
\end{align}
The components of the extrinsic curvature are given by
\begin{align}
 &\bar{K}^\mu{}_\nu = K^\mu{}_\nu = \fr{2\alpha}g^{\mu\alpha}( \partial_\rho g_{\alpha \nu} - \nabla_\alpha \beta_\nu - \nabla_\nu \beta_\alpha),\\
 & \bar{K}^a{}_b = K_\Sigma\, \delta^a{}_b :=\fr{\alpha}(\partial_\rho - \beta^\mu \partial_\mu) \log h\, \delta^a{}_b,\\
 &\bar{K}^a{}_\mu = 0,\quad \bar{K}^\mu{}_a = \cA_a \bar{K}^\mu{}_\psi - \delta^\mu{}_\psi A_b \bar{K}^b{}_a,
\end{align}
where the quantities without the bar are for $g_{\mu\nu}$.
The mean curvature is written as
\begin{align}
 \bar{K} = K + (2N-2) K_\Sigma .
\end{align}
Eq.~(\ref{eq:adm-evol0}) are decomposed to
\begin{subequations}\label{eq:evols}
\begin{align}
\fr{\alpha} \partial_\rho K^\mu{}_\nu = \bar{R}^\mu{}_\nu-\bar{K} K ^{\mu}{}_\nu 
+ \fr{\alpha} \cL_\beta K^\mu{}_\nu- \fr{\alpha}\nabla^\mu \nabla_\nu \alpha,
\end{align}
and
\begin{align}
\fr{\alpha} (\partial_\rho-\beta^\mu \partial_\mu) K_\Sigma = R_{\Sigma} - \bar{K} K_\Sigma 
- \fr{\alpha} \partial^\mu \alpha \partial_\mu \log h,
\end{align}
\end{subequations}
where the curvature tensors of the intrinsic metric are given by
\begin{align}
 \bar{R}^\mu{}_\nu = R^\mu{}_\nu - 2(N-1) (h^{-1} \nabla^\mu \nabla_\nu h - h^{-4} \delta^\mu{}_\psi g_{\nu \psi}).
\end{align}
and
\begin{align}
\bar{R}^a{}_b = R_{\Sigma} \delta^a{}_b := \left[2N h^{-2}-2 h^{-4} g_{\psi\psi} - (2N-3) h^{-2} (\nabla^\mu h) (\nabla_\mu h)-h^{-1} \nabla^2 h\right]\delta^a{}_b.
\end{align}

The nonzero components of the constraints are given by
\begin{align}
\bar{\nabla}_{\bar{\nu}} \bar{K}^{\bar{\nu}}{}_\mu - \bar{\nabla}_\mu \bar{K} 
= \nabla_\nu K^\nu{}_\mu - \nabla_\mu \bar{K} + 2(N-1) \partial_\nu \log h K^\nu{}_\mu
- 2N \partial_\mu \log h K_\Sigma =0,
\end{align}
and
\begin{align}
\bar{K}^2 - K^\mu{}_\nu K^\nu{}_\mu - (2N-2) K_\Sigma^2- R =0.
\end{align}
To solve the metric~(\ref{eq:ansatz}) in the $1/D$-expansion, we set
\begin{align}
\rho = \sR := r^{2N},\quad \beta_\mu dx^\mu := \frac{U}{2N\sR^{1-\fr{2N}}} e^{(0)},\quad \alpha := \sqrt{-\beta_\mu \beta^\mu},
\end{align}
and
\begin{align}
g_{\mu\nu} \xi^\mu \xi^\nu = -A (e^{(0)})^2 - 2 C_i  e^{(0)} e^{(i)} + H_{ij} e^{(i)}e^{(j)},\quad h = r \sin\theta. 
\end{align}

\section{Asymptotic behavior in the boosted ansatz}\label{eq:asym-bc}
In this section, we relate the asymptotic form of the ansatz~(\ref{eq:ansatz}) with the flat background
\begin{align}
ds^2 = - d\bar{t}^2 +dr^2 + r^2 \bar{\Phi}^2+r^2 \sin^2\theta \cos^2\theta\Psi^2+r^2d\theta^2+r^2 \sin^2\theta d\Sigma^2.
\end{align}
where $\bar{\Phi}:=d\bar{\phi}+\sin^2\theta \Psi$.
We assume the horizon is placed around $r=r_0$. To keep the calculation explicit, we do not set $r_0=1$ throughout this section. One can easily obtain the formula used in the main part by setting $r_0=1$.
First, we switch to the Eddington-Finkelstein coordinate,
\begin{align}
ds^2 = - dt^2 + r^2 \Phi^2 + \frac{2 (dt-r_0^2\Omega \Phi) dr}{\sqrt{1-\frac{r_0^4 \Omega^2}{r^2}}}
+r^2 \sin^2\theta \cos^2\theta\Psi^2+r^2d\theta^2+r^2 \sin^2\theta d\Sigma^2,\label{eq:app-bg-ef}
\end{align}
by the transformation
\begin{align}
 d\bar{t} = dt - \frac{dr}{\sqrt{1-\frac{r_0^4 \Omega^2}{r^2}}},\quad d\bar{\phi} = d\phi  - \frac{r_0^2 \Omega}{r^2}  \frac{dr}{\sqrt{1-\frac{r_0^4 \Omega^2}{r^2}}},
\end{align}
where $\Phi := d\phi+\sin^2\theta \Psi$.
In terms of the local boosted frame around $r=r_0$, 
\begin{align}
 e^{(0)} := \gamma (dt- r_0^2 \Omega \Phi),\quad e^{(1)} :=r_0 \gamma(\Phi - \Omega dt),\quad e^{(2)} := r_0 d\theta,\quad e^{(3)} := r_0 \sin\theta \cos\theta \Psi,
\end{align}
where $\gamma:=(1-r_0^2 \Omega^2)^{-1/2}$, the background geometry~(\ref{eq:app-bg-ef}) is written as
\begin{align}
&ds^2 = - \gamma^2(1-r^2\Omega^2)(e^{(0)})^2 +  \frac{2e^{(0)} dr}{\gamma \sqrt{1-\frac{r_0^4 \Omega^2}{r^2}}} + 2 r_0 \Omega \gamma^2 \left(\frac{r^2}{r_0^2}-1\right)e^{(0)}e^{(1)} \nonum
&\quad +\frac{r^2}{r_0^2}\left[\gamma^2\left(1-\frac{r_0^4\Omega^2}{r^2}\right) (e^{(1)})^2 +(e^{(2)})^2+(e^{(3)})^2\right].
\end{align}
With $\sR :=(r/r_0)^{2N}$, this determines the asymptotic behavior of the metric functions in
\begin{align}
 ds^2 = -A (e^{(0)})^2+2U e^{(0)} dr - 2C_i e^{(0)}e^{(i)} + H_{ij} e^{(i)}e^{(j)}+r^2 \sin^2 \theta d\Sigma^2,
\end{align}
as
\begin{align}
 A = \gamma^2 (1-r^2 \Omega^2) + \ord{\sR^{-1}} = 1 - \frac{ r_0^2\Omega^2}{1-r_0^2\Omega^2} \frac{\log \sR}{N}+\ord{N^{-2},\sR^{-1}},
\end{align}
\begin{align}
 U = \sqrt{\frac{1-r_0^2\Omega^2}{1-\frac{r_0^4\Omega^2}{r^2}}} + \ord{\sR^{-1}}
 = 1 + \frac{r_0^2\Omega^2}{1-r_0^2 \Omega^2} \frac{\log \sR}{N} + \ord{N^{-2},\sR^{-1}},
\end{align}
\begin{align}
C_i = \frac{r_0 \Omega \left(1-\frac{r^2}{r_0^2}\right)}{1-r_0^2\Omega^2} \delta_{i1}+\ord{\sR^{-1}} = - \frac{ r_0 \Omega}{1-r_0^2 \Omega^2}  \delta_{i1}\frac{\log \sR}{N} + \ord{N^{-2},\sR^{-1}}, 
\end{align}
and
\begin{align}
 H_{ij} =\frac{r^2}{r_0^2} \left [\delta_{ij} + \frac{r_0^2 \Omega^2\left(1-\frac{r_0^2}{r^2}\right)}{1-r_0^2\Omega^2}\delta_{i1}\delta_{j1}\right]+\ord{\sR^{-1}}=\delta_{ij} + \left(\delta_{ij}+\frac{ r_0^2\Omega^2\delta_{i1}\delta_{j1}}{1-r_0^2\Omega^2}\right)\frac{\log \sR}{N}+ \ord{N^{-2},\sR^{-1}}.
\end{align}

\section{Local event horizon, entropy and temperature}\label{sec:eventhorizon}
In this section, we derive the formula involving the local event horizon in the metric~(\ref{eq:ansatz}) without the large $D$ limit.
The local event horizon $r=r_h(t,\phi,\theta)$ is defined as a null hypersurface $||dr-dr_h||^2=0$~\cite{Bhattacharyya:2008xc}.
 The derivative of $r_h$ is written as
\begin{align}
dr_h = \partial_0 r_h e^{(0)} + \partial_i r_h e^{(i)} ,
\end{align}
where the dual basis is 
\begin{align}
 \partial_0 := \gamma ( \partial_t + \Omega \partial_\phi),\quad
  \partial_1 := \gamma ( \partial_\phi+  \Omega \partial_t),\quad
  \partial_2 := \partial_\theta,\quad \partial_3 := \csc\theta \sec\theta \partial_\psi - \tan \theta \partial_\phi.
\end{align}
The null condition becomes
\begin{align}
\left. A - 2 U \partial_0 r_h +  H_{ij}v^i v^j
\right|_{H}=0,
\end{align}
where
\begin{align}
 v^i :=\left. H^{ij} (C_j - U \partial_i r_h)\right|_H.
\end{align}
The horizon cross section is then given by
\begin{align}
ds_H^2 =  H_{ij} ( e^{(i)}-v^i e^{(0)})(e^{(j)}-v^j e^{(0)})+ r_h^2 \sin^2 \theta
 d\Sigma^2.
\end{align}
In the coordinate basis, this is written as
\begin{align}
ds_H^2 =\tilde{H}_{ab} ( dz^a - v^a dt)(dz^b-v^b dt) +r_h^2 \sin^2\theta d\Sigma^2.
\end{align}
where $a,b=\phi,\theta,\psi$ and
\begin{align}
v^\phi =  \frac{\Omega+v^1}{1+\Omega v^1}-v^\psi \sin^2 \theta, \quad v^\theta= v^2 \frac{\gamma(1-\Omega^2)}{1+\Omega v^1},
\quad v^\psi = \frac{v^3}{\cos \theta \sin\theta} \frac{\gamma(1-\Omega^2)}{1+\Omega v^1}.
\end{align}
It is also convenient to define
\begin{align}
v^\Phi := v^{\phi} + v^\psi \sin^2\theta = \frac{\Omega+v^1}{1+\Omega v^1}.
\end{align}
$H_{ij}$ and $\tilde{H}_{ab}$ are related by
\begin{align}
\left( \begin{array}{c} e^{(1)}-v^1 e^{(0)}\\
 e^{(2)}-v^2 e^{(0)}\\
 e^{(3)}-v^3 e^{(0)} 
 \end{array}\right)
 = \left(\begin{array}{ccc} \gamma (1-\Omega v^\Phi)&0&0\\
 \gamma \Omega v^2&1&0\\
 \gamma \Omega v^3 &0&\sin\theta \cos\theta \end{array} \right)
\left( \begin{array}{c} \Phi-v^\Phi dt,\\
 d\theta - v^\theta dt \\
 \Psi - v^\psi dt
 \end{array}\right),\label{eq:frametrans}
\end{align}
which leads to
\begin{align}
 \sqrt{{\rm det} \tilde{H}} = \sin\theta \cos\theta\gamma (1-\Omega v^\Phi) \sqrt{{\rm det} H}.
\end{align}
Thus, the entropy formula is given by
\begin{align}
 S = \frac{n\Omega_{n+1}}{4G\gamma} \int_0^{\pi/2} d\theta \int_0^{2\pi} \frac{d\phi}{2\pi} \frac{ r_h^{n-1} \cos \theta \sin^{n-1} \theta}{1+\Omega v^1} \sqrt{{\rm det}H},\label{eq:def-ent}
\end{align}
where we used the property
\begin{align}
 1- \Omega v^\Phi = \frac{1-\Omega^2}{1+\Omega v^1}.
\end{align}

On the other hand, the null generator of the horizon becomes
\begin{align}
 \xi = \partial_t + v^\phi \partial_\phi + v^\theta \partial_\theta + v^\psi\partial_\psi.
\end{align}
If $\xi$ is the Killing vector on the horizon, one can find the surface gravity is given by
\begin{align}
  \kappa =\left. \frac{\partial_r \tilde{A}}{2  U}\frac{1}{\gamma(1+\Omega v^1)}
  \right|_H,
\end{align}
where
\begin{align}
 \tilde{A} = A - 2 U \partial_0 r_H + 2 v^i ( C_i-U\partial_i r_H),
\end{align}
and hence the Hawking-Bekenstein temperature
\begin{align}
T = \frac{\kappa}{2\pi} =\left. \frac{\partial_r \tilde{A}}{4\pi  U}\frac{1}{\gamma(1+\Omega v^1)}\right|_H.\label{eq:def-temp}
\end{align}

\section{$D=2N+3$ Myers-Perry black holes}
In this section, we review the properties of Myers-Perry solution 
in $D=2N+3$, whose metric, in general, is given by~\cite{Myers:1986un}
\begin{align}
 ds^2 = - dt^2 + \frac{\mu \bar{r}^2}{\Pi F} \left(dt + \sum_{i=0}^N a_i \mu_i^2 d\phi_i\right)^2+\frac{\Pi F}{\Pi-\mu \bar{r}^2} d\bar{r}^2
 + \sum_{i=0}^N (\bar{r}^2+a_i^2)(d\mu_i^2+\mu_i^2 d\phi_i^2),\label{eq:mpodd-original}
\end{align}
where $\mu_i$ is the directional cosine which satisfies $\sum_{i=0}^N \mu_i^2=1$ and
\begin{align}
F=1-\sum_{i=0}^N \frac{a_i^2 \mu_i^2}{\bar{r}^2+a_i^2},\quad \Pi = \prod_{i=0}^N(\bar{r}^2+a_i^2).
\end{align}

We particularly interested in the case where the majority of spins are equal.

\subsection{Equal angular momenta : $N+1$ equal spins}\label{sec:exactmp-n+1-spin}
First, we revisit the known expression with equal spins $a_i=a$~\cite{Kunduri:2006qa}. In the new coordinates
\begin{align}
r=\sqrt{\bar{r}^2+a^2}, \quad \phi=\phi_0,\quad\zeta^\alpha = (\mu_\alpha/\mu_0) e^{i(\phi_\alpha-\phi_0)},
\end{align}
where $\{\zeta^\alpha\}_{\alpha=1,\dots,N}$ are the Fubini-Study coordinates for $CP^N$,
the metric~(\ref{eq:mpodd-original}) reduces to
\begin{align}
ds^2 = -\frac{F(r)}{H(r)}dt^2+\frac{dr^2}{F(r)}+r^2 H(r)^2(d\phi+\cA_N+\Omega(r)dt)^2+r^2 d\Sigma^2_N,
\end{align}
where $\cA_N$ and $d\Sigma_N^2$ is the K\"ahler potential and Fubini-Study metric on $CP^{N}$.
With $\mu:=r_0^{2N}$, the metric functions are written by 
\begin{align}
F(r) = 1 - \frac{r_0^{2N}}{r^{2N}}+ \frac{a^2 r_0^{2N}}{r^{2N+2}},\quad
 H(r) = 1 + \frac{a^2 r_0^{2N}}{r^{2N+2}},\quad
 \Omega(r)=-\frac{a r_0^{2N}}{r^{2N+2} H(r)}.
\end{align}
The position of the event horizon is determined by
\begin{align}
F(r_+) = 1 - \frac{r_0^{2N}}{r_+^{2N}}+ \frac{a^2 r_0^{2N}}{r_+^{2N+2}}=0.
\end{align}
The thermodynamic variables for this solution are obtained as follows
\begin{align}
&M = \frac{\Omega_{2N+1}}{8\pi G} \mu \left(N+\fr{2}\right),\quad J_\phi = \frac{\Omega_{2n+1}}{8\pi G} \mu (N+1)a,\\
&\Omega_\phi = \frac{a}{r_+^2},\quad S = \frac{\Omega_{2N+1}}{4G} \mu^{1/2} r_+^{N+1},\quad
 T = \frac{\kappa}{2\pi}= \frac{N \mu^{1/2}}{2\pi r_+^{N+1}}\left(1-\frac{N+1}{N}\frac{a^2}{r_+^2}\right).
\end{align}
With the decomposition~(\ref{eq:CPNtoCPNm1}), one can also calculate
\begin{align}
 J_{\psi} = \frac{N}{N+1} J_\phi,\quad \Omega_\psi =0.
\end{align}
In particular, the mass normalized angular momentum and entropy are given by
\begin{align}
& j_\phi := \frac{8\pi G J_\phi }{(N+1)\Omega_{2N+1}} \left(\frac{16\pi G M}{(2N+1)\Omega_{2N+1}}\right)^{-\frac{2N+1}{2N}}=\frac{a}{r_0},\nonum
& j_\psi := \frac{8\pi G J_\psi }{N\Omega_{2N+1}} \left(\frac{16\pi G M}{(2N+1)\Omega_{2N+1}}\right)^{-\frac{2N+1}{2N}}=\frac{a}{r_0},\nonum
& s := \frac{4G S}{\Omega_{2N+1}} \left(\frac{16\pi G M}{(2N+1)\Omega_{2N+1}}\right)^{-\frac{2N+1}{2N}}=\frac{r_+}{r_0} \sqrt{1-\frac{a^2}{r_+^2}}.
\end{align}

\subsection{Almost equal angular momenta : $N$ equal spins} \label{sec:exactmp-n-spin}
Now, we assume $N$ of $N+1$ spins are the same
\begin{align}
a_0 = a, \quad a_i = b \ (i=1,\dots,N).
\end{align}
We introduce the following coordinates
\begin{align}
&r=\sqrt{\bar{r}^2+b^2},\quad \theta = \arccos \mu_0 ,\nonum
&  \phi = \phi_0, \ \psi = \phi_N-\phi_0, \ u^i = (\mu_i/\mu_N) e^{i (\phi_i-\phi_N)},
\end{align}
where $\{u^i\}_{i=1,\dots,N-1}$ are the coordinates in the Fubini-Study metric for $CP^{N-1}$.
With this, the metric~(\ref{eq:mpodd-original}) reduces to
\begin{align}
&ds^2 = - \frac{F\Delta}{H}dt^2+\frac{G}{F}dr^2+r^2 G d\theta^2
 + r^2 \sin^2\theta \left(1+\frac{b^2\mu \sin^2\theta}{r^{2N+2}G}\right)(d\psi+\cA+V_\psi dt)^2\nonum
 &\quad +2  \left(r^2+\frac{\mu b(a+(b-a)\sin^2\theta)}{r^{2N}G}\right)\sin^2\theta (d\phi+V_\phi dt)(d\psi+\cA+V_\psi dt)\nonum
 &\quad +\left(r^2+(a^2-b^2)\cos^2\theta + \frac{\mu(a+(b-a)\sin^2\theta)}{r^{2N} G}\right)(d\phi+V_\phi dt)^2+r^2 \sin^2\theta d\Sigma^2,
\end{align}
where $\cA$ and $d\Sigma^2$ are the K\"ahler potential and the metric on $CP^{N-1}$, respectively.
The functions in the metric are given as follows
\begin{align}
& F = 1 +\frac{a^2-b^2}{r^2} -\frac{\mu (r^2-b^2)}{r^{2N+2}},\quad G = 1+\fr{r^2}(a^2-b^2)\sin^2\theta,\nonum
&H = 1 + \frac{a^2-b^2}{r^2} + \frac{\mu a^2 }{r^{2N+2}},\quad B = H + \frac{(a^2-b^2)F}{r^2} \sin^2\theta,\nonum
&V_\phi = \frac{a\mu}{r^{2N+2}B},\quad V_\psi=\frac{(a-b)(b^2+ab-r^2)\mu}{r^{2N+4}B},\nonum
& \Delta = 1 + \frac{(r^2+a^2-b^2)(a^2-b^2) \mu \sin^2\theta}{r^{2N+4}B}.
\end{align}
One can easily see that this metric reduces to the equally-rotating case by setting $a=b$ and using the decomposition in eq.~(\ref{eq:CPNtoCPNm1}).
The horizon is given by the largest root of 
\begin{align}
 F(r_+) = 1 + \frac{a^2-b^2}{r_+^2} - \frac{\mu(r_+^2-b^2)}{r_+^{2N+2}} =0.
\end{align}
Then, one can write $a$ as function of $r_+$ and $b$.
The thermodynamics are given by
\begin{align}
& M = \frac{\Omega_{2N+1}\mu}{8\pi G}\left(N+\fr{2}\right) ,\quad J_\phi =  \frac{\Omega_{2N+1}\mu}{8\pi G} (a+N b) , \quad   J_\psi = \frac{\Omega_{2N+1}\mu}{8\pi G} N b, \nonum
& \Omega_{\phi} = \frac{a}{a^2-b^2+r_+^2},\quad \Omega_{\psi} = \frac{b}{r_+^2} -  \frac{a}{a^2-b^2+r_+^2},\quad S = \frac{ \Omega_{2N+1}\mu}{4 G} \sqrt{r_+^2-b^2},\nonum
&   T = \frac{\kappa}{2\pi} = \frac{1}{2\pi \sqrt{r_+^2-b^2}}\left(N\left(1-\frac{b^2}{r_+^2}\right)-\frac{a^2}{r_+^2+a^2-b^2}\right). 
\end{align}
It is easy to check the Smarr formula and first law hold by differentiating with $\mu, r_+$ and $b$,
\begin{align}
&\frac{2N}{2N+1} M = T S + \Omega_\phi J_\phi + \Omega_\psi J_\psi ,\\
& \delta M = T \delta S + \Omega_\phi \delta J_\phi + \Omega_\psi \delta J_\psi.
\end{align}
With $\mu :=r_0^{2N}$, the scale invariant expressions are given as
\begin{subequations}\label{eq:mpthermonormal}
\begin{align}
 &j_\phi = \frac{8\pi G J_\phi}{(N+1)\Omega_{2N+1}} \left(\frac{8\pi G M}{(N+1/2)\Omega_{2N+1}}\right)^{-1-\fr{2N}}= \frac{Nb+a}{(N+1)r_0},\\
& j_\psi = \frac{8\pi G J_\psi}{N\Omega_{2N+1}} \left(\frac{8\pi G M}{(N+1/2)\Omega_{2N+1}}\right)^{-1-\fr{2N}}= \frac{b}{r_0},\\
 &s = \frac{4 GS}{\Omega_{2N+1}} \left(\frac{8\pi G M}{(N+1/2)\Omega_{2N+1}}\right)^{-1-\fr{2N}} = \frac{r_+}{r_0}\sqrt{1-\frac{b^2}{r_+^2}},
\end{align}
and
\begin{align}
\tilde{\Omega}_\phi = r_0 \Omega_\phi = \frac{a r_0}{a^2-b^2+r_+^2},\quad
\tilde{\Omega}_\psi = r_0 \Omega_\psi = \frac{br_0}{r_+^2}-\frac{a r_0}{a^2-b^2+r_+^2}.
\end{align}
\end{subequations}

\paragraph{Large $D$ limit}
To compare with the solution in the main part, we take the large $D$ limit ( or large $N$ limit ) of the thermodynamic variables.
Since $r_+ = r_0 + \ord{N^{-1}}$, we obtain
\begin{align}
 \tilde{\Omega}:=  \tilde{\Omega}_\phi + \tilde{\Omega}_\psi \simeq \frac{b}{r_0},\quad  \tilde{\Omega}_\psi \simeq \frac{b}{r_0}-\frac{a r_0}{a^2-b^2+r_0^2}\label{eq:exactmp-om-ompsi}
\end{align}
and
\begin{align}
 j_\phi \simeq j_\psi \simeq \frac{b}{r_0} \simeq \tilde{\Omega} ,\quad s \simeq \sqrt{1-\frac{b^2}{r_0^2}} \simeq \sqrt{1-\tilde{\Omega}^2}.
 \label{eq:exactmp-ds}
\end{align}
The difference between two angular momenta becomes
\begin{align}
\Delta j := 2N(j_\psi-j_\phi) \simeq \frac{2(b-a)}{r_0}.\label{eq:exactmp-dj}
\end{align}
Since eq.~(\ref{eq:exactmp-om-ompsi}) leads to
\begin{align}
 \frac{a}{r_0} \simeq \frac{1\pm\sqrt{1-4 (1-\tilde{\Omega} ^2) (\tilde{\Omega} -\tilde{\Omega}_\psi )^2}}{2 (\tilde{\Omega}  -\tilde{\Omega}_\psi )},
\end{align}
eq.~(\ref{eq:exactmp-dj}) can be expressed as the relation between the normalized quantities
\begin{align}
\Delta j \simeq  2\tilde{\Omega} - \frac{1\pm\sqrt{1-4 (1-\tilde{\Omega} ^2) (\tilde{\Omega} -\tilde{\Omega}_\psi )^2}}{\tilde{\Omega}  -\tilde{\Omega}_\psi }.\label{eq:exactmp-dj1}
\end{align}

%
%

\end{document}